%% file: g2light.tex
\documentclass[aps,prd,superscriptaddress,amsmath,twocolumn,nofootinbib,superscriptaddress,10pt]{revtex4}

\newcommand{\amul}{599(6)(8)\times10^{-10}}
\newcommand{\amutot}{667(6)(12)\times10^{-10}}

\newcommand{\amuHVP}{a_\mu^{\scriptscriptstyle \mathrm{HVP,LO}}}
\usepackage{color,hhline,psfrag,rotating,amssymb}
\usepackage{dcolumn}
\usepackage{verbatim}
\usepackage{bm}

\newcommand{\ctitle}{\multicolumn{1}{c}}
\begin{document}
\title{The hadronic vacuum polarization contribution to $a_{\mu}$ from full lattice QCD}

\author{Bipasha Chakraborty}
\affiliation{SUPA, School of Physics and Astronomy, University of Glasgow, Glasgow, G12 8QQ, UK}
\author{C.~T.~H.~Davies}
\email[]{christine.davies@glasgow.ac.uk}
\affiliation{SUPA, School of Physics and Astronomy, University of Glasgow, Glasgow, G12 8QQ, UK}
\author{P.~G. de Oliveira}
\affiliation{SUPA, School of Physics and Astronomy, University of Glasgow, Glasgow, G12 8QQ, UK}
\author{J.~Koponen}
\affiliation{SUPA, School of Physics and Astronomy, University of Glasgow, Glasgow, G12 8QQ, UK}
\author{G.~P. Lepage}
\affiliation{Laboratory for Elementary-Particle Physics, Cornell University, Ithaca, New York 14853, USA}
\collaboration{HPQCD collaboration}
\homepage{http://www.physics.gla.ac.uk/HPQCD}
\noaffiliation
\author{R.~S. Van de Water}
\affiliation{Fermi National Accelerator Laboratory, Batavia, IL, USA}

\date{\today}

\begin{abstract}
We determine the contribution to the anomalous magnetic moment
of the muon from the $\alpha^2_{\mathrm{QED}}$ hadronic
vacuum polarization diagram using full lattice QCD
and including $u/d$ quarks with physical masses for the first time.
We use gluon field configurations that include $u$, $d$, $s$
and $c$ quarks in the sea at multiple values of the lattice spacing,
multiple $u/d$ masses and multiple volumes that allow us
to include an analysis of finite-volume effects.
We obtain a result for
$\amuHVP$ of $\amutot$, where the first error is
from the lattice calculation and the second includes systematic
errors from missing QED and isospin-breaking effects and from
quark-line disconnected diagrams.
Our result implies a discrepancy between the experimental determination
of $a_{\mu}$ and the Standard Model of 3$\sigma$.
\end{abstract}


\maketitle

\section{Introduction}
The muon's gyromagnetic ratio~$g_\mu$ is known experimentally with
extremely high accuracy: its magnetic anomaly, $a_\mu\equiv(g_\mu-2)/2$,
has been measured to~0.5\,ppm~\cite{Bennett:2006fi} and
a new experiment aims to reduce that uncertainty to~0.14\,ppm~\cite{Venanzoni:2012yp}.
By comparing these results with Standard Model predictions, we can use the
muon's anomaly to search for indirect evidence of new physics beyond the
mass range directly accessible at the Large Hadron Collider. There are tantalizing
hints of a discrepancy between theory and experiment\,---\,the difference
is currently~2.2(7)\,ppm~\cite{snowmass}\,---\,but more
precision is needed. In particular the Standard Model prediction, which
currently is known to about~0.4\,ppm~\cite{snowmass},
must be substantially improved in order to match the expected improvement
from experiment.

\begin{figure}
\centering
\includegraphics[width=0.3\textwidth]{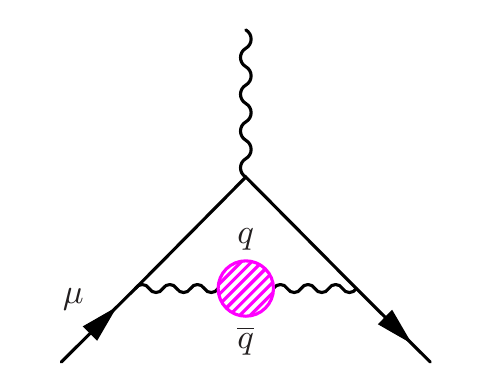}
\caption{
The $\alpha^2_{\mathrm QED}$ hadronic vacuum polarization contribution to the
muon anomalous magnetic moment is represented as a shaded
blob inserted into the photon propagator (represented
by a wavy line) that corrects the point-like
photon-muon coupling at the top of the diagram.
}
\label{fig:hvp}
\end{figure}

The largest theoretical uncertainty in~$a_\mu$ comes from the
vacuum polarization of hadronic matter (quarks and gluons) as illustrated
in Figure~\ref{fig:hvp}. This contribution has been estimated to
a little better than~1\% (which is 0.6\,ppm of $a_\mu$)
from experimental data on $e^+e^- \rightarrow$ hadrons and $\tau$
decay~\cite{Davier, Hagiwara:2011af, Jegerlehner:2011ti, Benayoun:2015gxa, Jegerlehner:2015stw},
but much recent
work~\cite{aubin-blum, renner, boyle, wittig, Burger:2013jya, Chakraborty:2014mwa, Francis:2014qta, Malak:2015sla, Gregory:2015bno, Spraggs:2016jcx}
has focused on a completely different approach, using
Monte Carlo simulations of lattice QCD~\cite{Blum:2013qu}, which
promises to deliver smaller errors in the future.

In an earlier
paper~\cite{Chakraborty:2014mwa}, we introduced a new technique for the
lattice QCD analyses that allowed us to calculate
the $s$~quark's vacuum-polarization contribution from
Figure~\ref{fig:hvp} with
a precision of~1\% for the first time.
Here we extend that analysis to the much more important
(and difficult to analyze)
case of~$u$ and~$d$ quarks, allowing
us to obtain the complete contribution from hadronic vacuum polarization at $\alpha^2_{QED}$.
We achieve a precision of~2\%, for the first time from lattice QCD.
A large part of our uncertainty is from
QED, isospin breaking and quark-line disconnected effects that were not
included in the simulations, but will be in future simulations. The
remaining systematic errors add up to only~1\%. A detailed
analysis of these systematic errors allows us to map out a strategy for
reducing lattice QCD errors well below~1\% using computing resources that
are substantial but currently available.

\begin{table*}
\caption{
Here we use gluon field configurations from the MILC
collaboration~\cite{Bazavov:2010ru, milchisq}.
$\beta=10/g^2$ is the QCD gauge coupling, and $w_0/a$~\cite{fkpi}
gives the lattice spacing, $a$, in terms of the Wilson flow parameter,
$w_0$~\cite{Borsanyi:2012zs}.
We take $w_0$=0.1715(9)\,fm fixed from $f_{\pi}$~\cite{fkpi}.
The lattice spacings are approximately 0.15\,fm for sets~1--3,
0.12\,fm for sets~4--8, and 0.09\,fm for sets~9--10.
$L$ and $T$
are the spatial and temporal dimensions of the lattice.
$am_\ell, am_s$ and $am_c$ are the masses in lattice units
of light ($m_\ell\equiv m_u=m_d$),
strange, and charm quarks in the sea, with $am_s^{\mathrm{phys}}$
giving the correct $m_s$ value on that ensemble~\cite{Chakraborty:2014aca}.
Valence quark masses equal $m_\ell$ except for set~4, where
$am_\ell^\mathrm{val}=0.01044$ is slightly different from $am_\ell$.
$am_{\pi}$ and $am_\rho$ give the corresponding
masses for the $\pi$~and $\rho$~mesons; $f_\rho$ is the lattice
result for the $\rho$'s leptonic decay constant.
$Z_{V,\overline{s}s}$ gives the vector
current renormalization factor (calculated for $s$ quarks)
obtained nonperturbatively~\cite{Chakraborty:2014zma}.
The number of
configurations is given
in the final column; we use 16 time sources on each and average over the
spatial polarizations for the vector current. We tested for autocorrelations
by binning  configurations, but found no effect.
}
\label{tab:params}
\begin{ruledtabular}
\begin{tabular}{cclllllllclcr}
Set & $\beta$ &
\multicolumn{1}{c}{$w_0/a$} &
\multicolumn{1}{c}{$am_\ell$} &
\multicolumn{1}{c}{$am_s$} &
\multicolumn{1}{c}{$am_c$} &
\multicolumn{1}{c}{$am_s^\mathrm{phys}$} &
\multicolumn{1}{c}{$am_{\pi}$} &
\multicolumn{1}{c}{$am_{\rho}$} &
\multicolumn{1}{c}{$f_\rho/(m_\rho Z_{V,\overline{s}s})$} &
\multicolumn{1}{c}{$Z_{V,\overline{s}s}$} &
{$L/a\times T/a$} &
\multicolumn{1}{c}{$n_{\mathrm{cfg}}$} \\
\hline
1 & 5.8 & 1.1119(10) & 0.013 & 0.065 & 0.838 & 0.0700(9)   & 0.23643(9)   & 0.6679(15) & 0.2659(9)  & 0.9887(20) & $16\times48$ & 9947\\
2 & 5.8 & 1.1272(7) & 0.0064 & 0.064 & 0.828 & 0.0686(8)   & 0.16617(7)   & 0.6128(47) & 0.2677(19)  & 0.9887(20) & $24\times48$ & 1000\\
3 & 5.8 & 1.1367(5) & 0.00235 & 0.0647 & 0.831 & 0.0677(8) & 0.10172(4) & 0.5968(45) & 0.2776(16)  & 0.9887(20) & $32\times48$ & 997\\
                                                                                                       \hline
4 & 6.0 & 1.3826(11) & 0.0102 & 0.0509 & 0.635 & 0.0545(7) & 0.18938(8) & 0.5276(35) & 0.2635(23)  & 0.9938(17) & $24\times64$ & 1053\\
5 & 6.0 & 1.4029(9) & 0.00507 & 0.0507 & 0.628 & 0.0533(7) & 0.13492(8) & 0.4938(82) & 0.2625(63)  & 0.9938(17) & $24\times64$ & 1020\\
6 & 6.0 & 1.4029(9) & 0.00507 & 0.0507 & 0.628 & 0.0533(7) & 0.13415(5) & 0.4866(49) & 0.2635(34)  & 0.9938(17) & $32\times64$ & 1000\\
7 & 6.0 & 1.4029(9) & 0.00507 & 0.0507 & 0.628 & 0.0534(7) & 0.13401(6) & 0.4850(46) & 0.2652(31)  & 0.9938(17) & $40\times64$ & 331\\
8 & 6.0 & 1.4149(6) & 0.00184 & 0.0507 & 0.628 & 0.0527(6)  & 0.08162(4) & 0.4730(27) & 0.2771(11)  & 0.9938(17) & $48\times64$ & 998\\
                                                                                                           \hline
9 & 6.3 & 1.9006(20) & 0.0074 & 0.037 & 0.44 & 0.0378(5)    & 0.14062(10)    & 0.3854(37) & 0.2626(29)  & 0.9944(10) & $32\times96$ & 1000\\
10 & 6.3 & 1.9330(20) & 0.00363 & 0.0363 & 0.43 & 0.0366(5) & 0.09850(10) & 0.3508(42) & 0.2683(33) & 0.9944(10) & $48\times96$ & 298\\
\end{tabular}
\end{ruledtabular}
\end{table*}

\section{Lattice QCD Calculation}
Almost all of the hadronic vacuum polarization contribution (HVP) comes from
\emph{connected} diagrams with the structure shown in Figure~\ref{fig:hvp}:
 the photon creates a quark and antiquark which propagate, while interacting
with each other, and eventually annihilate back into a photon. Here we
analyze the case where the photon creates either a $u\bar u$ or $d\bar d$
pair; we calculated contributions from heavier quarks
in~\cite{Chakraborty:2014mwa, Donald:2012ga, Colquhoun:2014ica}.
Disconnected diagrams, where the quarks and
antiquarks created by the photons annihilate into gluons rather than photons,
give much smaller contributions~\cite{Chakraborty:2015ugp, Blum:2015you};
we will discuss these at the end of this paper.

In  Section~\ref{sec:extracting-amu},
we describe how we extract $a_\mu$ from a single
configuration set. Unlike in our previous analysis with
$s$~quarks~\cite{Chakraborty:2014mwa}, the light-quark
vacuum polarization becomes
very noisy at large~$t$ for physical masses. We introduce a simple
procedure for improving the signal-to-noise ratio in this calculation.

In Section~\ref{sec:correction-amu}, we examine the largest systematic
errors in our lattice analysis. These come
from finite-volume effects, and, more importantly,
from mass splittings between
different tastes of pion in our HISQ formalism. We address these errors
in two ways.

First we use chiral perturbation theory to calculate corrections, including
contributions from the leading term and the largest
corrections to it. We also calculate contributions from a variety of other
higher-order corrections in order to assess their impact on~$a_\mu$.

The second way in which we address our systematic errors is to extract
values for~$a_\mu$ from simulations with much larger light-quark
masses\,---\,approximately 2.5~and 5~times the physical mass\,---\,where
systematic errors from finite volumes and staggered pions
become negligible. As discussed in~\cite{Burger:2013jya},
most of the light-quark mass dependence of $a_\mu$ can be removed by
rescaling the vacuum polarization with appropriate powers
of~$m_\rho^\mathrm{latt}/m_\rho^\mathrm{expt}$. Here we show that rescaled
results from large masses are consistent
with the corrected results from physical masses,
giving us confidence in both types of result. We combine
all of our results into a single global fit from which we
extract a final result.

\subsection{Extracting $a_\mu$}
\label{sec:extracting-amu}
The leading-order contribution to the muon anomalous magnetic moment
from the HVP is obtained
by inserting the quark vacuum polarization into the
photon propagator~\cite{Blum:2002ii,Lautrup:1971jf}.
Ignoring disconnected contributions,
the vacuum polarization separates
into distinct contributions for each quark flavour, $\mathrm{f}$:
\begin{equation}
\amuHVP(\mathrm{f}) =
\frac{\alpha}{\pi}
\int_0^{\infty} dk^2 \,f(k^2)
\left(4\pi\alpha \right)
\hat{\Pi}_{\mathrm{f}}(k^2)
\label{eq:amu}
\end{equation}
where $\alpha\equiv\alpha_{\mathrm{QED}}$ is the QED fine structure constant
and $k$~is the (Euclidean) momentum carried
by the virtual photons. $f(k^2)$ is a kinematic factor that diverges
as $k^2 \rightarrow 0$, where the
renormalized vacuum polarization function,
$\hat\Pi(k^2)\equiv\Pi(k^2)-\Pi(0)$,
vanishes.
The resulting integrand is peaked around $k^2 \approx m^2_{\mu}$.
Note that $\hat\Pi(k^2)$ includes a factor of
$Q_{\mathrm{f}}^2$, where $Q_{\mathrm{f}}$ is
the electric charge of quark $\mathrm{f}$ in units
of the proton's charge. This is a change to the convention
that we used in~\cite{Chakraborty:2014mwa}.

Lattice QCD is used to calculate the vacuum polarization function~$\hat\Pi(k^2)$.
In~\cite{Chakraborty:2014mwa} we developed an accurate method
for evaluating Eq.~(\ref{eq:amu}) by
defining~$\hat\Pi(k^2)$ in terms of its Taylor expansion,
\begin{equation}
\hat\Pi(k^2) = \sum_{j=1}^\infty k^{2j} \,\Pi_j,
\label{eq:pihat}
\end{equation}
where the Taylor coefficients $\Pi_j$
are determined from time-moments~$G_{2j}$
of the vector current-current correlator at zero spatial momentum:
\begin{align}
    G_{2j} &\equiv
    \sum_t \sum_{\vec{x}} t^{2j} Z_V^2 \langle j^{i}(\vec{x},t)j^{i}(0) \rangle  \nonumber \\
    Q^2_{\mathrm{f}}G_{2j} &= (-1)^j \left. \frac{\partial^{2j}}{\partial k^{2j}} k^2\hat{\Pi}(k^2) \right|_{k^2=0}
    \nonumber \\
    &= (-1)^j \,(2j)!\,\, \Pi_{j-1}.
\label{eq:G}
\end{align}
Here $Z_V$ renormalizes the lattice vector current, and
\begin{equation}
t \in (0, 1, 2, \ldots T/2-1,0,-T/2+1, \ldots, -2,-1) .
\end{equation}
We replace the Taylor series by its $[n,n]$ and $[n,n-1]$ Pad\'e
approximants for the integral in Eq.~(\ref{eq:amu}). The approximants
provide an accurate approximation for both the
low and high~$k^2$ regions in the integral, and results converge
to better than~1\% by~$n=2$, with the exact result bracketed by results from
the $[n,n]$ and $[n,n-1]$  approximants for
each~$n$~\cite{Chakraborty:2014mwa}. We evaluate
the integral numerically.

Signal/noise in lattice QCD (Monte Carlo)
evaluations of vector correlators degrades
exponentially as the time separation, $t$, between source and sink
increases. This increases the uncertainties in
the $\Pi_j$ from~Eq.(\ref{eq:G}), especially
as~$j$ increases.
The noise problem is particularly
acute for correlators made of $u/d$ quarks, because
the $\rho$ (which controls
the signal) is much more massive than the $\pi$ (which
controls the noise)~\cite{Lepage:1989hd}.
The values that a correlator can take at large $t$, however, are
constrained by its values at smaller $t$ and the known form
of the correlator.
Thus we reduce the noise in our Taylor coefficients by replacing
the correlator at large-$t$ with its value determined
from a standard multi-exponential fit (to data at all~$t$s, large
and small).
We use
\begin{equation}
\label{eq:tstar}
G(t) = \left\{ \begin{array}{cc} G_{\mathrm{data}}(t), \quad & t\le t^{\ast}  \\
        {G}_{\mathrm{fit}}(t), \quad  & t > t^{\ast}\end{array}\right. .
\end{equation}
and test that our results are stable on varying $t^{\ast}$.  
We find $a_\mu$ to be independent of $t^{\ast}$
to better than $\pm0.5\%$ for $t^\ast$~values ranging between~0.5\,fm
and~1.5\,fm (our default value since larger values
lead to larger statistical errors). 
To further improve
our results, we calculate a $2\times2$ matrix of vector correlators
that combines the local operator we need for the time-moments
(Eq.~(\ref{eq:G})) with a
smeared operator that overlaps more strongly with the ground-state
vector meson (the $\rho$).
Using the fit
at large~$t$ also allows us to correct for the
finite temporal length of the lattice. 

We present more details on this noise-reduction strategy in Appendix~\ref{appendix:corr}.
There we show that this strategy introduces a new uncertainty into our analysis,
due to low-energy ($<m_\rho$) $\pi\pi$~states in the simulations. We
estimate this uncertainty using chiral perturbation theory. We also show 
that the uncertainty is bounded by the variation of $a_\mu$ as $t^\ast$~is 
changed from~0.5\,fm to~1.5\,fm. Our estimate is consistent with 
the variation in~$a_\mu$ mentioned above, so we include an
uncertainty of~ $\pm0.5$\% in our error budget to allow for these effects.

\begin{figure}
\centering
\includegraphics[width=0.5\textwidth]{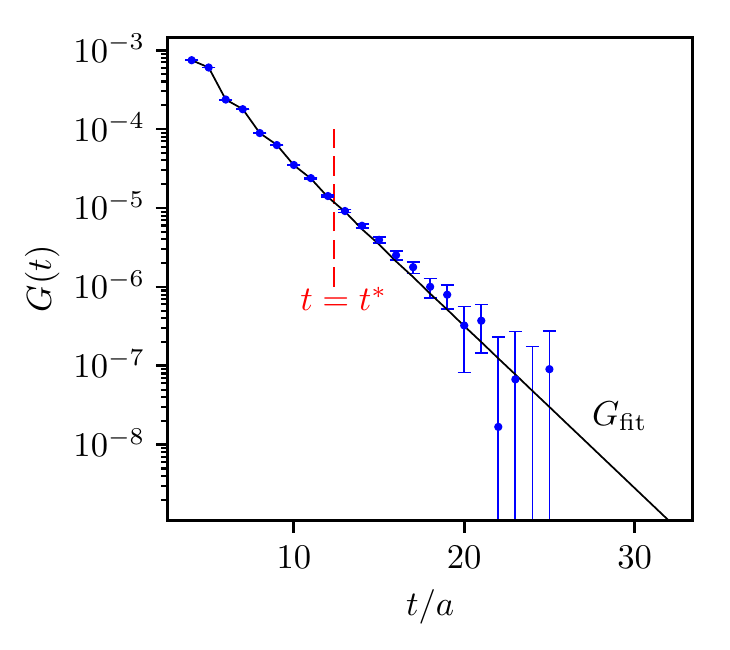}
\caption{
Monte Carlo data for the current-current correlator on configuration set~8
from Table~\ref{tab:data} compared with the fit function~$G_\mathrm{fit}$
that replaces the
data for $t>t^{\ast}$.
$G_\mathrm{fit}$ is obtained from a multi-exponential fit to
all the data shown, above and below~$t^*$, together with additional
data for correlators with smeared sources. Its uncertainty
is of order the width of the line
at large~$t$, and smaller at small~$t$.
The oscillations at small~$t$ are an artifact of staggered quarks whose
contribution to $a_\mu$~is small and vanishes with the lattice spacing.
}
\label{fig:tstar}
\end{figure}

\newcommand{\cPi}[1]{\hfill $\Pi_#1$ \hfill\mbox{}}
\begin{table*}
    \caption{Columns 2-5 give the uncorrected Taylor coefficients $\Pi_j$ (Eq.~\ref{eq:pihat}),
    in units of $1/\mathrm{GeV}^{2j}$,
    for each of the lattice data sets in Table~\ref{tab:params}.
    The errors given include statistics and the (correlated) uncertainty from setting the
    lattice spacing using $w_0$; the latter error largely cancels in our analysis.
    Estimates of the connected contribution from $ud$-quarks to $\amuHVP$
    are given for each of the $[1,0]$, $[1,1]$, $[2,1]$ and $[2,2]$
    Pad\'e approximants in columns 6-9; results are multiplied by~$10^{10}$.
    These estimates are obtained after correcting the moments, as discussed
    in the text. The final estimate for~$\amuHVP$ is given in the last column.}
    \label{tab:data}
    \begin{ruledtabular}
        \begin{tabular}{clllllllll}
Set & \ctitle{$\Pi_1$} &
\ctitle{$\Pi_2$} &
\ctitle{$\Pi_3$} &
\ctitle{$\Pi_4$} &
\ctitle{$[1,0]\times10^{10}$} &
\ctitle{$[1,1]\times10^{10}$} &
\ctitle{$[2,1]\times10^{10}$} &
\ctitle{$[2,2]\times10^{10}$} &
\ctitle{$\amuHVP\times10^{10}$}\\
\hline
\hline
1 & $0.0624\hfill(7)$  & $-0.0760\hfill(17)$ & $0.102\hfill(3)$ & $-0.138\hfill(6)$ & $660.1\hfill(3.2)$ & $590.7\hfill(2.8)$ & $593.3\hfill(2.9)$ & $592.2\hfill(2.9)$ & $592.7\ \,(3.0)$\\
2 & $0.0729\hfill(11)$ & $-0.1028\hfill(31)$ & $0.159\hfill(8)$ & $-0.250\hfill(16)$ & $663.8\hfill(6.0)$ & $591.7\hfill(5.2)$ & $595.8\hfill(5.5)$ & $593.6\hfill(5.5)$ & $594.7\ \,(5.6)$\\
3 & $0.0796\hfill(13)$ & $-0.1182\hfill(39)$ & $0.190\hfill(10)$ & $-0.311\hfill(21)$ & $689.9\hfill(9.2)$ & $604.4\hfill(8.9)$ & $618.3\hfill(11.3)$ & $609.1\hfill(10.5)$ & $613.7(11.5)$\\
\hline
4 & $0.0638\hfill(8)$  & $-0.0803\hfill(21)$ & $0.111\hfill(5)$ & $-0.157\hfill(9)$ & $650.0\hfill(5.7)$ & $582.5\hfill(4.9)$ & $585.0\hfill(4.9)$ & $584.0\hfill(4.9)$ & $584.5\ \,(4.9)$\\
5 & $0.0715\hfill(13)$ & $-0.0992\hfill(41)$ & $0.151\hfill(11)$ & $-0.236\hfill(24)$ & $653.4\hfill(14.1)$ & $583.4\hfill(11.8)$ & $587.4\hfill(12.0)$ & $585.3\hfill(11.9)$ & $586.4(11.9)$\\
6 & $0.0736\hfill(11)$ & $-0.1052\hfill(33)$ & $0.166\hfill(9)$ & $-0.267\hfill(19)$ & $650.8\hfill(8.3)$ & $581.5\hfill(7.0)$ & $585.4\hfill(7.2)$ & $583.4\hfill(7.1)$ & $584.4\ \,(7.2)$\\
7 & $0.0744\hfill(11)$ & $-0.1075\hfill(34)$ & $0.171\hfill(9)$ & $-0.277\hfill(20)$ & $652.9\hfill(7.8)$ & $583.5\hfill(6.6)$ & $587.2\hfill(6.8)$ & $585.2\hfill(6.7)$ & $586.2\ \,(6.8)$\\
8 & $0.0811\hfill(12)$ & $-0.1239\hfill(36)$ & $0.206\hfill(9)$ & $-0.345\hfill(21)$ & $675.1\hfill(7.6)$ & $593.6\hfill(7.5)$ & $606.9\hfill(9.6)$ & $597.9\hfill(8.9)$ & $602.4(10.0)$\\
\hline
9 & $0.0625\hfill(9)$ & $-0.0778\hfill(25)$ & $0.107\hfill(6)$ & $-0.151\hfill(11)$ & $640.1\hfill(7.3)$ & $574.2\hfill(6.2)$ & $576.6\hfill(6.2)$ & $575.7\hfill(6.2)$ & $576.2\ \,(6.2)$\\
10 & $0.0755\hfill(13)$ & $-0.1109\hfill(41)$ & $0.178\hfill(11)$ & $-0.292\hfill(25)$ & $652.1\hfill(8.4)$ & $583.4\hfill(7.2)$ & $586.7\hfill(7.3)$ & $585.0\hfill(7.2)$ & $585.8\ \,(7.2)$\\
    \end{tabular}
    \end{ruledtabular}
\end{table*}

We work on ensembles of gluon field configurations that have
an improved discretization of
the gluon action~\cite{Hart:2008sq} and sea $u$, $d$, $s$, and $c$ quarks using the
Highly Improved Staggered Quark (HISQ) action~\cite{HISQ_PRD,HISQ_PRL}.
They were generated by the MILC collaboration~\cite{milchisq}.
We have results for three lattice spacings,
for $u/d$~quark masses ranging from~$m_s/5$
down to the physical value, and for three lattice volumes,
for one combination of masses and lattice spacing.
These results allow us to test and correct for
the most important systematic errors in our simulations.
We approximate $m_u = m_d \equiv m_\ell$, using the same
masses for valence and sea quarks.
The ensembles are described in Table~\ref{tab:params}.

A by-product of the fits to our lattice QCD correlators are values
for the $\rho$~mass and decay constant for a variety of $u/d$~quark masses.
Our results agree with experiment to within errors
for realistic quark masses (see Appendix~\ref{appendix:corr}).
This is an important test of the
correlators we use to calculate~$\amuHVP$.

The local vector current that we use is not the conserved vector current
for this quark action and so must be renormalized. We do this
nonperturbatively by demanding that the vector form factor for this current be 1 between two
equal mass mesons at rest ($q^2=0$)~\cite{Chakraborty:2014zma}.
We use pseudoscalar mesons made of $s$ quarks to do this, on
the $m_\ell/m_s$=0.2 ensembles at each lattice spacing and give values
in Table~\ref{tab:params}. We ignore the mass dependence of the $Z_V$s
since it is less than~$0.1\%$ ($\mathcal{O}((am_s/\pi)^2\alpha_s )$)
and therefore negligible compared to our statistical errors.

\subsection{Correcting~$a_\mu$}
\label{sec:correction-amu}
The Taylor coefficients $\Pi_j$ from each of
our $u/d$-quark vector correlators are listed in Table~\ref{tab:data}.
We introduce two corrections, one after the other, before calculating $\amuHVP$
in order to minimize our systematic errors:
\begin{enumerate}
\item \emph{Reduce lattice artifacts:} We correct our
results for errors caused by the finite
spatial volume of the lattice and by artifacts from using
(HISQ) staggered quarks (mass splittings between
pions of different taste).
We do this with an effective theory, derived from chiral perturbation theory,
that couples $\rho$s, $\pi^+\pi^-$~pairs, and $\gamma$s. We use the theory
to calculate the $\Pi_j$
for both the continuum and our lattice QCD calculation, and we correct
our lattice QCD moments with the differences.
The corrections for each moment and
configuration set are given in Table~\ref{tab:stagg-pipi}
of Appendix~\ref{appendix:effth}. The largest
corrections for $a_\mu$ are for our lightest pions and turn out to be
around $+7$\%. They are an order of magnitude
smaller for our heaviest pions. These corrections lead to an
uncertainty of $\pm0.7\%$ in our final~$a_\mu$.

Chiral perturbation theory is well suited to our analysis because
our moments are calculated at $q^2=0$, where chiral perturbation
theory is valid.
The dominant finite-volume and staggered-pion corrections
come from leading-order pion vacuum
polarization, $\gamma\to\pi^+\pi^-\to\gamma$,
as discussed in~\cite{Aubin:2015rzx}. This correction
can be calculated quite accurately because it is determined by
the (well measured) charge and mass of the pion. We find that it
is five times larger than the other corrections.
The next largest contribution comes from corrections to  the
$\gamma$--$\pi\pi$~vertex due to the pion's charge radius. We
include both of these corrections in our final result, together
with a variety of the other
higher-order corrections that allow us to explore the rate of
convergence of chiral perturbation theory.
See Appendix~\ref{appendix:effth} for more details.

\item \emph{Reduce $m_\ell$ dependence}: We rescale $m_\ell$ to its physical
value
in the $\pi\pi$ and $\rho$ contributions to $a_\mu$ (80\% of the total)
to reduce $a_\mu$'s strong dependence on~$m_\ell$. We do this
in three steps, modifying and extending a method introduced
in~\cite{Burger:2013jya}:

\noindent (a) We remove the
vacuum polarization contribution to $a_{\mu}$ due
to $\gamma\to\pi^+\pi^-\to\gamma$
using the continuum effective field theory with the pion mass set equal to the
simulation result for the Goldstone pion.

\noindent (b) We rescale the resulting $\Pi_j$
by $(m_{\rho}^{\mathrm{latt}}/m_{\rho}^{\mathrm{expt}})^{2j}$.
This reduces $m_\ell$~dependence because the $\rho$~meson pole dominates the
vacuum polarization, especially once the $\pi^+\pi^-$ contribution
is removed.
We also find that rescaling removes further finite volume
dependence; see the end of Appendix~\ref{appendix:effth}.
Rescaling has a large impact for our heavier-than-physical
$m_\ell$ values, but has little effect
for physical $m_\ell$ where the simulation's $m_{\rho}$ agrees with experiment.
We apply the Pad\'e approximants at this stage to generate estimates
for $\amuHVP$, since they converge more quickly without the $\pi^+\pi^-$
contribution (which is restored in the next step).

\noindent (c) We reintroduce the $\pi^+\pi^-$ contribution removed
in step~2a, but with
the pion mass set equal to $m_{\pi^+}$ (139.6\,MeV)
rather than the pion masses from the simulations.
Again this has little impact for
our configurations with physical $m_\ell$~values.
\end{enumerate}
Our results, using different
Pad\'e approximants, are shown in Table~\ref{tab:data}. Our
final result for each configuration is obtained by taking
a value half way between results from the $[2,2]$ and $[2,1]$~Pad\'e
approximants, with an associated error equal to half the difference~\cite{Chakraborty:2014mwa}.
Our statistics do not permit the use of higher-order approximants.

\begin{figure}
\centering
\includegraphics[width=0.5\textwidth]{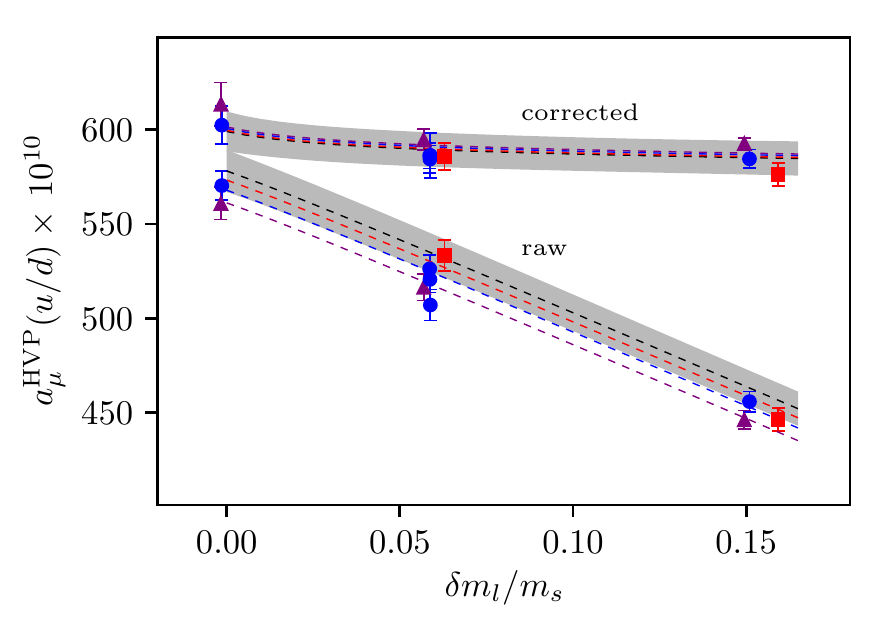}
\caption{
Our results for the connected $u/d$ contribution to $\amuHVP$
as a function of the $u/d$ quark mass (expressed as its deviation from
the physical value in units of the physical $s$ quark mass).
The lower curve shows our uncorrected data; the upper
curve includes correction factors discussed in the text and is
used to obtain the final result.
Data come
from simulations with lattice spacings of 0.15\,fm
(purple triangles), 0.12\,fm (blue circles), and
0.09\,fm (red squares). The gray bands show the $\pm1\sigma$~predictions
of our model (Eq.~(\ref{eq:amu-fit})) after fitting it to the data.
The dashed lines show the results from the fitting function for each
lattice spacing (colored as above) and extrapolated to zero lattice
spacing (black).
The $\chi^2$ per degree of freedom was~1.0 and~0.6 for the upper and lower
fits, respectively.
}
\label{fig:amu}
\end{figure}

Our corrected results are plotted in Figure~\ref{fig:amu}, together
with the results without corrections (labeled ``raw''). The corrected
results are nearly independent of~$m_\ell$, as expected. Residual dependence
comes from other hadronic channels in the vacuum polarization beyond
the $\pi^+\pi^-$ and $\rho$ contributions. The corrected results also
show smaller $a^2$ and volume~dependence, as is
particularly clear from the points for $\delta m_\ell/m_s$~just above~0.05.

The final step in our analysis is to fit the corrected results from our 10~ensembles
to a function of the form
\begin{align}
\amuHVP 
\left( 1 +
c_\ell \frac{\delta m_\ell}{\Lambda}
+ c_s \frac{\delta m_s}{\Lambda}
+ \tilde c_\ell \frac{\delta m_\ell}{m_\ell}
+ c_{a^2}\frac{(a\Lambda)^2}{\pi^2}
\right)
\label{eq:amu-fitfunction}
\end{align}
where $\delta m_\mathrm{f} \equiv m_\mathrm{f} - m_\mathrm{f}^\mathrm{phys}$,
and $\Lambda\equiv5m_s$ is of order the QCD scale~(0.5\,GeV). The fit
parameters have the following priors:
\begin{equation}
    c_\ell = 0(1)\quad  c_s = 0.0(3)
    \quad  \tilde c_\ell = 0.00(3)\quad c_{a^2} = 0(1)
    \label{eq:amu-fit}
\end{equation}
together with prior~$600(200)\times10^{-10}$ for~$\amuHVP$.
This fit corrects for mis-tuned quark masses and the finite lattice spacing.
The first two correction
terms
allow for residual dependence on~$m_\ell$ and (slight) mistuning
in the $s$~quark's mass. We expect smaller corrections from the latter
because it enters only through the quark sea.
The last term in Eq.~(\ref{eq:amu-fitfunction}) corrects for
the finite-lattice spacing. Note that our
analysis is quite insensitive to uncertainties in the lattice
spacing because the leading dependence on the lattice spacing cancels
when we rescale our moments with the lattice result for the
rho mass (step~2 in our analysis).

The $\delta m_\ell/m_\ell$~correction in Eq.~(\ref{eq:amu-fitfunction})
is associated with steps~2(a) and~2(c) in our analysis,
where we replace the (continuum) $\gamma\to\pi\pi\to\gamma$
contribution to~$a_\mu$ corresponding to the simulation's pion mass
with the same contribution evaluated at the physical pion mass.
The $\Pi_1$ Taylor coefficient dominates~$a_\mu$ (see the $[1,0]$ entries
in Table~\ref{tab:data})
and therefore Eq.~(\ref{eq:pipi-moment})
in Appendix~\ref{appendix:finvol} implies
that $a_\mu$'s dependence on the light-quark
mass~$m_\ell$ is given approximately by
\begin{equation}
    \label{eq:ml-dependence}
    a_\mu(m_\ell) \approx a_0 \left(
    1 + d_\ell \frac{m_\ell^\mathrm{phys}}{m_\ell}
    \right)
\end{equation}
where $a_0$ and $d_l$ depend weakly on~$m_\ell$, $m_\ell^\mathrm{phys}$
is the physical value for $m_\ell$, and the physical value
for~$a_\mu$ is approximately $a_0(1+d_\ell)$.
The bulk of the $d_\ell$~term comes
from $\gamma\to\pi\pi\to\gamma$ vacuum polarization, with $d_\ell\approx0.1$.
So steps~2(a) and~2(c) in our analysis procedure have the effect of replacing
$m_\ell$ by $m_\ell^\mathrm{phys}$, thereby bringing $a_\mu(m_\ell)$ closer to
its physical value. About a quarter of $d_\ell$ comes from sources other
than the simple $\pi\pi$ vacuum polarization\,---\,the most important is
from $\gamma\to\rho\to\pi\pi\to\gamma$.  Thus our analysis steps~2(a) and~2(c)
do not fully correct the $m_\ell$ in Eq.~(\ref{eq:ml-dependence}). There
is a residual piece of order $a_0 \times 0.2\, d_\ell \,\delta m_\ell/m_\ell$ that
we account for with the $\delta m_\ell/m_\ell$~correction in our fit formula.
In practice the contribution from this term is comparable to
our statistical errors, and so has marginal impact on our final result.

We tested our fit by adding higher-order terms in the various corrections
and cross terms. None of these variations changed our final results by more than
a small fraction of the final uncertainty.

We also tested our fit by dropping various configuration sets. Dropping
the configuration sets with the heaviest pions (sets 1, 4 and 9) shifts
our final result for~$a_\mu$ by less than a fifth of a standard deviation and
leaves the total error unchanged. Dropping the sets with physical pion masses
(sets 3 and 8) shifts the final result by a standard
deviation and increases the final error by~30\%. Each variation is consistent,
within errors, with the full analysis.

\begin{table}
\caption{Error budget for the connected contributions to
the muon anomaly~$a_\mu$ from vacuum polarization of $u/d$ quarks.}
\label{tab:errbudget}
\begin{ruledtabular}
\begin{tabular}{rc}
                                & $\amuHVP(u/d)$ \\
\hline
               QED corrections: &                   1.0\,\% \\
  Isospin breaking corrections: &                   1.0\,\% \\
Staggered pions, finite volume: &                   0.7\,\% \\
    Correlator fits ($t^\ast$): &                   0.5\,\% \\
        $m_\ell$ extrapolation: &                   0.4\,\% \\
        Monte Carlo statistics: &                   0.4\,\% \\
           Pad\'e approximants: &                   0.4\,\% \\
       $a^2\to0$ extrapolation: &                   0.2\,\% \\
             $Z_V$ uncertainty: &                   0.2\,\% \\
               Correlator fits: &                   0.2\,\% \\
       Tuning sea-quark masses: &                   0.2\,\% \\
   Lattice spacing uncertainty: &                   $<0.05$\,\% \\
\hline
                         Total: &                    1.8\,\% \\
\end{tabular}
\end{ruledtabular}
\end{table}

Our final result from the fit for the connected contribution from $u/d$~quarks
is $\amuHVP=\amul$, where the first error comes from the
lattice calculation and fit and the second
is due to missing contributions from QED and isospin breaking ($m_u\ne m_d$), each of
which we estimate to enter at the level of~1\% of the $u/d$ piece of
$\amuHVP$. These estimates are supported by more detailed
studies: The key isospin breaking effect
of $\rho-\omega$ mixing is estimated in~\cite{Wolfe:2010gf}
to make a $3.5\times 10^{-10}$
contribution (0.6\%) and the QED effect of producing a hadron
polarization bubble consisting
of $\pi^0$ and $\gamma$ is estimated in~\cite{Hagiwara:2003da} to make a $4.6\times 10^{-10}$ contribution (0.8\%). The leading contributions to
our final uncertainty are listed in Table~\ref{tab:errbudget}.
Note that our final result is~3.5\% above the extrapolated result from
the raw data shown in Fig.~\ref{fig:amu}; most of that shift comes
from corrections to the $\pi\pi$~vacuum polarization in chiral
perturbation theory.

\begin{figure}
\centering
\includegraphics[width=0.49\textwidth]{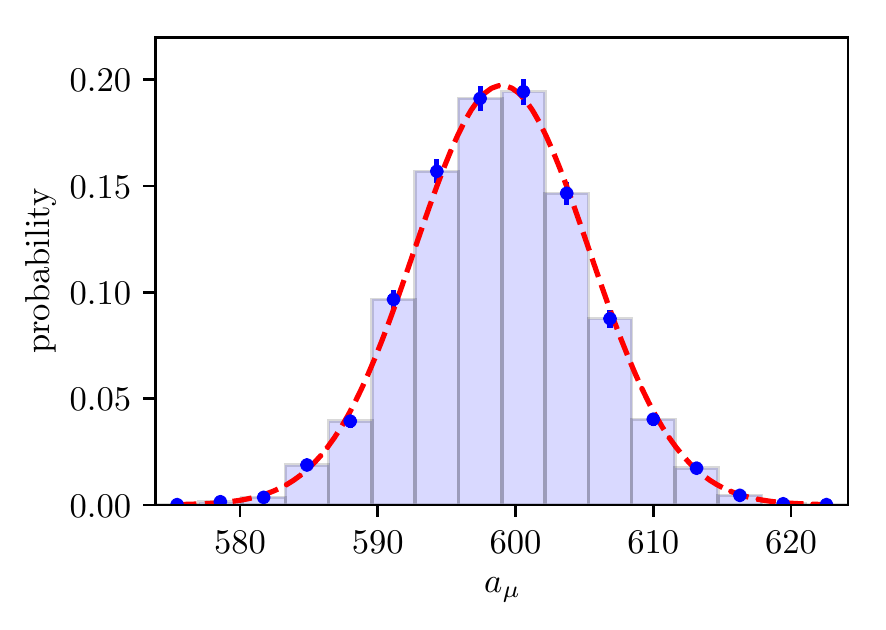}
\caption{ Bayesian probability distribution for~$\amuHVP(u/d)$ (bars)
compared with results from the least-squares fit (dashed line).
}
\label{fig:amu_bayes}
\end{figure}

We tested the validity of the least-squares fit that
determines our~$\amuHVP(u/d)$ by replacing the fit with a
Bayesian expectation value (a 16-dimensional numerical
integration) over the distributions of the input data
and priors. The results, in Fig.~\ref{fig:amu_bayes},
show that the least-squares fit (dashed-line)
agrees well with the probability distribution
from the corresponding Bayesian analysis~(bars).

\section{Discussion/Conclusions}

\begin{figure}
\centering
\includegraphics[width=0.48\textwidth]{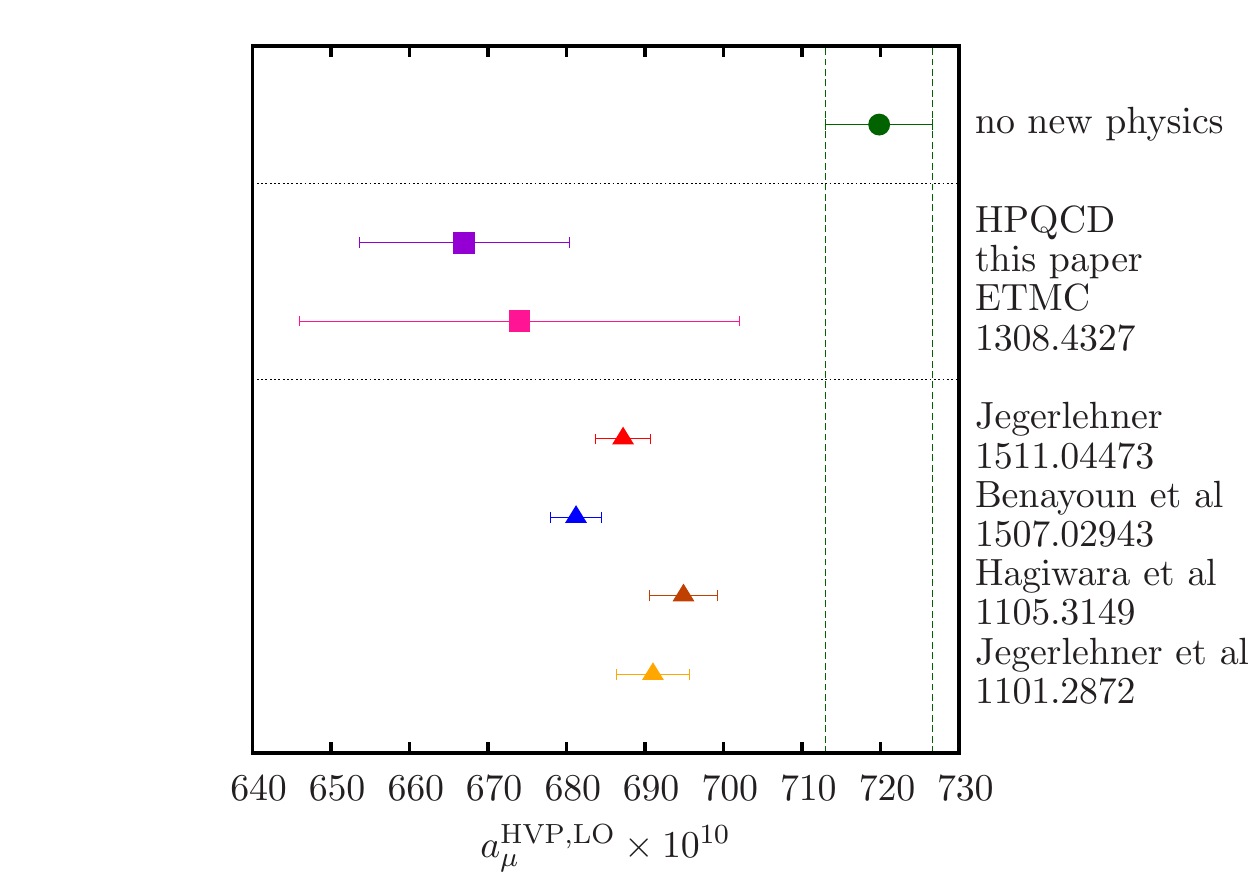}
\caption{
Our final result for $\amuHVP$ from lattice QCD
compared to an earlier lattice result (also with $u$, $d$, $s$ and $c$
quarks) from the ETM Collaboration~\cite{Burger:2013jya}, and to
recent results using experimental
cross-section
information~\cite{Jegerlehner:2015stw,Hagiwara:2011af,Jegerlehner:2011ti,Benayoun:2015gxa}.
We also compare with the result expected from the experimental value
for~$a_\mu$ assuming that there are no contributions from physics
beyond the Standard Model.
}
\label{fig:hvpcomp}
\end{figure}

Adding results from our earlier calculations for other quark
flavours~\cite{Chakraborty:2014mwa,Colquhoun:2014ica},
the connected contributions to $\amuHVP$ are:
\begin{equation}
    \left.\amuHVP\right|_\mathrm{conn.} \times 10^{10}
    =
    \begin{cases}
        599(11) & \mbox{from $u/d$ quarks} \\
        53.4(6) & \mbox{from $s$ quarks} \\
        14.4(4) & \mbox{from $c$ quarks} \\
        0.27(4) & \mbox{from $b$ quarks}
    \end{cases}
\end{equation}
We combine these results with our recent estimate~\cite{Chakraborty:2015ugp}
of the contribution
from disconnected diagrams involving $u$, $d$ and $s$ quarks, taking this
as~$0(9)\times10^{-10}$. This agrees with, but has a more conservative
uncertainty than, the value obtained in~\cite{Blum:2015you}.
We then obtain an estimate for the entire contribution
from hadronic vacuum polarization:
\begin{equation}
\amuHVP = \amutot
\label{eq:amufinal}
\end{equation}
This agrees well with the only earlier $u/d/s/c$ lattice QCD result,
$674(28)\times10^{-10}$~\cite{Burger:2013jya},
but has errors from the lattice calculation reduced by a factor of four.
It also agrees with earlier non-lattice results using
experimental data, ranging from ($\times 10^{10}$):
$694.9(4.3)$~\cite{Hagiwara:2011af}
to
$681.9(3.2)$~\cite{Benayoun:2015gxa}.
These
are separately more accurate than our result but have a spread
comparable to our uncertainty.
New results from BESIII~\cite{Ablikim:2015orh} may resolve this.

It is also useful to compare our result to the expectation from
experiment. Assuming there is no new physics beyond the Standard Model,
experiment requires $\amuHVP$ to be $720(7)\times 10^{-10}$. This value is
obtained by subtracting from experiment the accepted values of
QED~\cite{Aoyama:2012wk}, electroweak~\cite{Gnendiger:2013pva}, higher
order HVP~\cite{Hagiwara:2011af, Kurz:2014wya} and
hadronic light-by-light contributions~\cite{Prades:2009tw}:
\begin{eqnarray}
\label{eq:amu-nnp}
a_{\mu}^{\mathrm{HVP,LO,no\, new\, physics}} &=& a_{\mu}^{\mathrm{expt}}-a_{\mu}^{\mathrm{QED}}-a_{\mu}^{\mathrm{EW}} \nonumber \\
&-& a_{\mu}^{\mathrm{HVP,HO}} - a_{\mu}^{Hlbl}.
\end{eqnarray}

Figure~\ref{fig:hvpcomp} compares our results with others from
previous continuum and lattice analyses. We also compare with results
expected from experiment if there is no new physics contributing to~$a_\mu$.
The `no-new-physics' value is
roughly $3.5\sigma$ away from our result (Eq.~(\ref{eq:amufinal})), but
we need significantly smaller theoretical errors before we can make a case
for new physics.

From Table~\ref{tab:errbudget} we see that uncertainties
can be reduced by improving the calculation of the
quark-line disconnected contribution~\cite{Blum:2015you, Toth}
and from new simulations with $m_u \ne m_d$; this is straightforward.
Adding QED effects to a simulation is more difficult (see, for example,
\cite{Carrasco:2015xwa}), but it is particularly simple here
because the hadronic system is electrically neutral, so there are no
infrared divergences to be dealt with\footnote{There are higher order QED effects where
the photon interacts with both $\mu$ and hadrons (the `hadronic light-by-light'
contribution) which are more
complicated\,---\,lattice QCD also shows promise here~\cite{Blum:2015gfa}.}.

The remaining uncertainties are
together only about 1\%~of our answer. The largest (0.7\%) is caused
by $\alpha_sa^2$ differences in mass between pions of different
taste with HISQ quarks.
Reducing the lattice spacing
to~0.06\,fm at the physical pion mass would
cut this uncertainty in half. The remaining
errors would all be reduced by smaller lattice spacings and higher
statistics, both of which are feasible on time scales commensurate
with the schedule for the new experiments an $a_{\mu}$.

From our results we also obtain the total HVP contribution to the
electron: $a_e^{\mathrm{HVP,LO}}=0.01779(39) \times 10^{-10}$,
to be compared to $0.01846(12) \times 10^{-10}$ from $e^+e^-$
data~\cite{Jegerlehner:2015stw}.

{\it Acknowledgements.} We are grateful to the MILC collaboration
for the use of their
gauge configurations and code, to C. McNeile for updating
$w_0/a$ on set 9 and to C. DeTar,
G. Donald and T. Teubner for useful discussions.
Our calculations were done on the Darwin Supercomputer
as part of STFC's DiRAC facility jointly
funded by STFC, BIS
and the Universities of Cambridge and Glasgow.
This work was funded by the Gilmour bequest to the University of
Glasgow, the National Science Foundation,
the  Royal Society, STFC and the Wolfson Foundation.
Fermilab is operated by Fermi Research Alliance, LLC under 
contract number DE-AC02-07CH11359 with the U.S. Department of Energy.  

\appendix
\input{supplements}

\bibliography{g2light}

\end{document}

%% file: supplements.tex



\section{Correlator Fits}
\label{appendix:corr}
We construct a $2\times2$~matrix of meson propagators
using all combinations of two meson operators, with zero three momentum,
for the source and sink.
One meson operator (``loc'') is
the local vector current. The other (``sm'') is a vector current
but with smearing
applied to the quark field, using operator
\begin{equation}
\label{eq:smear}
\left[ 1+ \frac{r_0^2 D^2}{4n} \right]^n
\end{equation}
where $D^2$ is the covariant Laplacian operator and $r_0$ is a width parameter.
Since we are using staggered quarks and require current-current
correlators of a specific staggered taste, we use the {stride-2}
$D^2$ operator here, with the difference operator defined for grid
spacing~$2a$ (rather than~$a$). We choose $r_0=3a$, $3.75a$, and~$4.5a$
for lattices spacings 0.15\,fm, 0.12\,fm and~0.09\,fm, respectively,
with $n=20$, 30, and~40.

The result is a matrix of correlators, $G_{ij}$, where $i$ labels the source
and $j$ the sink. Each of $i,j$ is either
``loc'' for the local vector operator or ``sm''
for the smeared vector operator.
We fit $G_{ij}$ to the form:
\begin{align}
\label{eq:corrfit}
G_{ij}(t) &= a^3 \sum_{k=0}^{N-1} b_i^{(k)}\, b_j^{(k)}
\left( e^{-E^{(k)}t} + e^{-E^{(k)}(T-t)} \right)\\
&- (-1)^t
a^3 \sum_{k=0}^{N-1}
d_i^{(k)}\,d_j^{(k)}
\left( e^{-\tilde E^{(k)}t} + e^{-\tilde E^{(k)}(T-t)} \right)
\nonumber
\end{align}
where $k$ labels the energy eigenvalues appearing
in the correlator and $T$~is the temporal extent of the lattice.
The first sum is over $1^{--}$~vector states that couple
to the vector operators. The second
is over opposite-parity states that arise here because of our use of
staggered quarks; this term oscillates in sign as $t$~increases, which helps
the fit distinguish between it and the first term. We use a Bayesian approach
to the fitting~\cite{gplbayes} with the following fit parameters and
broad priors (in units of GeV):
\begin{align}
    \log(E^{(0)}) &= \log(0.75(38))
    \nonumber \\
    \log(E^{(k)} - E^{(k-1)}) &= \log(1.0(5))\quad(k>0)
    \nonumber \\
    \log(b_\mathrm{loc}^{(0)}) &=
    \begin{cases}
        \log(0.14(14)) & (k=0) \\
        \log(0.42(42)) & (k>0)
    \end{cases}
    \nonumber \\
    b_\mathrm{sm}^{(0)},\, b_\mathrm{sm}^{(k)} &= 0.01(1)
\end{align}
for the first sum, and the analogous parameters and priors for the
second sum but with
\begin{equation}
    \log(\tilde E^{(0)}) = \log(1.2(6)),
\end{equation}
to reflect the higher mass of the lowest opposite-parity state. To avoid
lattice artifacts (from the HISQ action) at very small times,
we fit the correlators only for $t$~values larger than 0.5--0.7\,fm.
We used~$N=5$, but get identical results with larger values of~$N$.
The fits were all excellent, with $\chi^2$ per degree of freedom ranging
between~0.6 and~1.1 in different fits.
The use of a smeared operator improves the fit results for
$E^{(0)}$ and $b_\mathrm{loc}^{(0)}$ (from which we obtain our values
for $m_\rho$ and $f_\rho$) by an amount
commensurate with its numerical cost.

\begin{figure}
\centering
\includegraphics[width=0.48\textwidth]{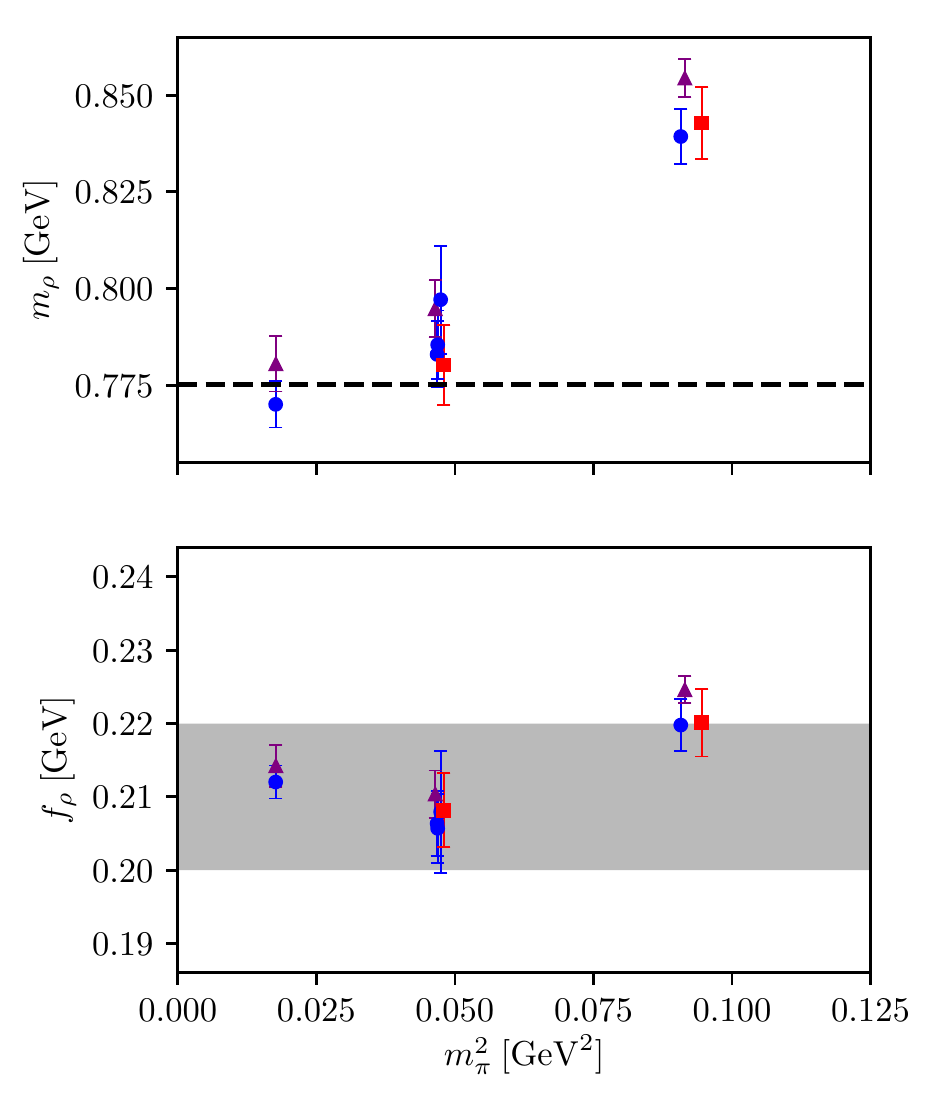}
\caption{
Results for the $\rho$ meson mass (upper plot)
and decay constant (lower plot) from the vector
correlators used to determine the $u/d$
connected contribution to $\amuHVP$. Results
are shown for different $u/d$ masses, as indicated
by the corresponding values of~$m_\pi^2$ (the
lightest being the physical value).
Data come
from simulations with lattice spacings of 0.15\,fm
(purple triangles), 0.12\,fm (blue circles), and
0.09\,fm (red squares). Experimental results
for the mass (dashed line) and decay constant
(gray band) are shown as well. A comparison
of our results with those of~\cite{boyle, Burger:2013jya}
is given in~\cite{Chakraborty:2015cso}.
}
\label{fig:rho}
\end{figure}

As discussed in the main text, we use a combination of data
and fit results when computing moments of the local current-current
correlator $G\equiv G_{\mathrm{loc},\mathrm{loc}}$:
\begin{equation}
    G(t) =
    \begin{cases}
        G_\mathrm{data}(t) & t\le t^\ast \\
        G_\mathrm{fit}(t) & t>t^\ast
    \end{cases}
\end{equation}
where we define
\begin{align}
G_\mathrm{fit}(t) &= a^3 \sum_{k=0}^{N-1} b_\mathrm{loc}^{(k)} \,b_\mathrm{loc}^{(k)}\,
 e^{-E^{(k)}t}\\
&- (-1)^t
a^3 \sum_{k=0}^{N-1}
\, d_\mathrm{loc}^{(k)}\, d_\mathrm{loc}^{(k)}\,
e^{-\tilde E^{(k)}t}
\nonumber
\end{align}
with the best-fit values for the parameters. $G_\mathrm{fit}$ is the
same as $G_{\mathrm{loc},\mathrm{loc}}$ from Eq.~(\ref{eq:corrfit})
but with~$T\to\infty$, thereby correcting for the
finite temporal extent of the lattice. Note that about 80\% of our
final result for~$a_\mu$ comes from $t\le t^\ast$ (=1.5\,fm),
where we use simulation data rather than our fit.

The sum over states in~$G_\mathrm{fit}$ (above) includes vector mesons like
the~$\rho$ and also multi-hadron states, which
enter as discrete energy eigenstates because of the finite spatial volume of our
lattice. The lowest-energy states are $\pi\pi$~states
for configurations with physical pion masses,
but we see no evidence of these in our fits\,---\,the dominant contribution
comes from the $\rho$~meson. This is expected because there are only a
few $\pi\pi$~states below the $\rho$~mass, and their contribution is suppressed
by a factor of one over the lattice volume (see Eq.~(\ref{eq:pipi-int}) below),
making their contributions to~$a_\mu$ smaller than our statistical errors.
Note that it has been possible to see coupled $\rho$ and $\pi\pi$ states
in lattice QCD calculations (see, for example,~\cite{Dudek:2012xn})
but to do so requires careful meson and multi-meson
operator optimization to achieve measurable overlaps; the calculations
do not use the local vector current that is relevant here.

The contribution of the low-energy $\pi\pi$~states coming from $t\le t^*$ is
included in our calculation, since we use the Monte Carlo results in that
region. The contribution from~$t>t^*$, however, is underestimated or missing.
That contribution can be calculated using the chiral
formalism developed below. We find that the low-energy $\pi\pi$~contribution
from $t>t^*$ should be $3\times10^{-10}$
when $t^*=1.5$\,fm (our default value), and so we 
include an uncertainty of $\pm3\times10^{-10}$ in 
our error budget for~$a_\mu$ to account for these states.
This estimate is for configuration set~8 in Table~\ref{tab:params}, where the
uncertainty is largest, so it is probably an overestimate of the impact on 
the entire calculation.

The $t>t^*$ contribution from low-energy $\pi\pi$~states
would be twice as large had we chosen $t^*=0.5$\,fm, and therefore
the difference
\begin{equation}
  \delta a_\mu \equiv a_\mu(t^*=0.5) - a_\mu(t^*=1.5)
\end{equation}
provides an upper bound on the possible error caused by
omitting these states from~$G_\mathrm{fit}$.
Redoing our full analysis for $t^*=0.5$\,fm, we find
that $\delta a_\mu=0\pm3\times10^{-10}$, which is consistent with our
direct estimate from chiral perturbation theory.

An important check on the quality of our correlators and fit is that the
$\rho$ mass and decay constant agree with experiment when the light quarks have their
physical values. This is illustrated by Figure~\ref{fig:rho}, which shows the
mass and decay constants from each of our configuration sets. Theory and
experiment agree to within errors for physical quark masses.\footnote{%
The definition of~$f_\rho$ is complicated by the large width of the
$\rho$~meson. Applying naive definitions gives results around~0.208\,GeV
from $\tau$~decay and around~0.220\,GeV from $\rho\to e\overline e$, with
errors of order a couple percent in each case. A more careful analysis,
which models non-resonant backgrounds in each case,
is needed to resolve the differences between these two channels. We take
the experimental value to be $f_\rho=0.21(1)$\,GeV for Figure~\ref{fig:rho}.}

\section{Finite Volume and Staggered Pions}
\label{appendix:finvol}
We use chiral perturbation theory to correct systematic errors in our
lattice results caused by the finite volumes of our
lattices, and by mass splittings between the different tastes of pion
generated by our (staggered-quark) HISQ discretization. Our general
strategy is to identify terms in the chiral expansion that are sensitive
to the volume and to  pion masses (i.e., loops with pions). We
calcuate these terms without and with lattice artifacts, and then
add the difference to the lattice results.

\begin{figure}
\centering
\includegraphics{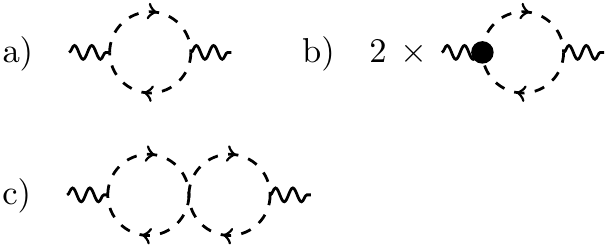}
\caption{Leading diagrams from chiral perturbation theory that contribute
to~$\delta\Pi_j$: a) leading-order $\pi^+\pi^-$ vacuum polarization;
b) vacuum polarization corrected for the pion's charge radius;
c) $\pi\pi$~scattering correction. Dashed lines represent pions.
}
\label{fig:chpth}
\end{figure}

The only relevant contribution from zeroth order in the chiral expansion
is the $\pi^+\pi^-$~vacuum polarization (Fig.~\ref{fig:chpth}a). As expected,
we find that it provides most of the correction.

There are three types of
higher-order correction beyond this term.
The first two involve corrections (from, for example, tadpole diagrams)
to the leading vacuum polarization
diagram that are suppressed by powers of
either the strange or the light-quark masses. Such corrections
are typically of order 10\% the leading contribution for $s$~quarks
and~1\% for $u/d$~quarks. We include a extra 10\%~uncertainty in our
corrections to account for such contributions.

The third type of correction involves terms suppressed
by powers of~$q^2/\Lambda^2$ where $\Lambda$ is the chiral
scale~($\approx1\,\mathrm{GeV}$).
Such terms are easily analyzed in our formalism, because it relies on
moments.
They enter in
first order as corrections to the $\gamma$-$\pi\pi$ vertex due to the pion's
charge radius (Fig.~\ref{fig:chpth}b).
$\pi\pi$~scattering (Fig.~\ref{fig:chpth}c)
also enters at this order, but is much less important,
as we shall see. Second-order and higher contributions
come from further corrections to the vertices, iterations of the leading
diagrams, and so on. These give small contributions compared with our errors.

Our analysis is simplified by using an extended version of standard chiral
perturbation theory that includes $\gamma$s, $\pi$s,
and~$\rho$s~\cite{Gasser:1983yg,Ecker:1988te}. Such a theory is particularly
useful here because $\gamma$-$\rho$ mixing accounts for the bulk of
the vacuum polarization
contribution to the muon anomaly. In the next section we examine
$\gamma$-$\rho^0$-$\pi^+\pi^-$ mixing to all orders in the leading
interactions that couple these channels. This analysis
includes all of the contributions illustrated in Fig.~\ref{fig:chpth},
as well as all iterations of these
diagrams.
It also includes an infinite number of $(q^2/m_\rho^2)^n$ corrections.

Following~\cite{Jegerlehner:2011ti}, we make one further simplification
in our analysis that
allows us, in effect, to absorb the 4-pion contact interaction into the
amplitude for $\pi\pi\to\rho\to\pi\pi$. This is done by replacing the
chiral $\rho$-$\pi\pi$ coupling by a simpler coupling,
$-ig_{\rho\pi\pi}(p+p^\prime)^\mu$, analogous to the photon coupling.
The resulting $\pi\pi$ scattering amplitude, which now comes entirely from
$\pi\pi\to\rho\to\pi\pi$, is equivalent
to what is obtained from
the chiral theory, but simpler to
analyze, at least for our application.

It is well known that chiral parameters for the $\rho$~channel in
$\pi\pi$~scattering are more or less saturated by the~$\rho$
itself~\cite{Ecker:1988te}. Our analysis relies upon this fact as
it uses high-order chiral coefficients determined by the~$\rho$.
After the pion charge and mass, the most important parameter for our
results is the pion charge radius. The
model we use below gives a pion charge radius squared
of~$0.46\,\mathrm{fm}^2$,
which agrees well with experiment
at~$0.45(1)\,\mathrm{fm}^2$~\cite{Olive:2016xmw}.
Similarly the $P$-wave scattering length
for $\pi\pi$ scattering in our model is $0.037/m_\pi^3$, which again
compares well with experiment at $0.038(2)/m_\pi^3$
(see, for example, \cite{GarciaMartin:2011cn} and
\cite{Pelaez:2004vs}.) These comparisons show that
the parameters in the effective theory
are tuned sufficiently well for our purposes.

In the next section we derive a photon propagator that
takes account of~$\gamma$-$\rho^0$-$\pi^+\pi^-$ mixing in our
effective field theory. We then specialize
that result for use in analyzing~$a_\mu$. Finally we
show how these results are affected by the lattice's finite
volume and by taste-splittings between HISQ pions. This allows us to
correct the moments from our simulation to remove systematic errors
from finite volumes and staggered pions.

\subsection{Photon propagator}

The one-loop corrected
$\rho$ and photon propagators of the effective theory,
\begin{equation}
    G(q) \equiv \left(
    \begin{matrix}
    G_{\gamma\gamma}(q) & G_{\gamma\rho}(q) \\
    G_{\gamma\rho}(q) & G_{\rho\rho}(q)
    \end{matrix}
    \right),
\end{equation}
are obtained by solving a matrix Lippmann-Schwinger equation,
\begin{equation}
    G = G_0  - G_0 \Sigma G,
\end{equation}
where
\begin{equation}
    \Sigma = q^2 \left(
    \begin{matrix}
        \Sigma_{\gamma\gamma}^{(1)} & e /g_\rho + \Sigma_{\gamma\rho}^{(1)} \\
        e/g_\rho + \Sigma_{\gamma\rho}^{(1)} & \Sigma_{\rho\rho}^{(1)}
    \end{matrix}
    \right).
\end{equation}
Here we project onto a transverse polarization to remove the spin algebra.
The lowest-order propagator is
\begin{equation}
    G_0 =
    {\left(
    \begin{matrix}
    q^2 & 0 \\
    0 & q^2 - m_{0\rho}^2
    \end{matrix}
    \right)}^{-1},
\end{equation}
while the leading-order $\pi\pi$ loops give amplitudes~\cite{Jegerlehner:2011ti}
\begin{align}
    \Sigma_{\gamma\gamma}^{(1)} &= e^2 \, \Sigma(q^2) \nonumber \\
    \quad \Sigma_{\gamma\rho}^{(1)} &= eg_{\rho\pi\pi}\,\Sigma(q^2)
    \quad\quad
    \quad \Sigma_{\rho\rho}^{(1)} = g_{\rho\pi\pi}^2 \,\Sigma(q^2)
\end{align}
where
\begin{align}
    48&\pi^2\,\Sigma(q^2) \nonumber \\
     &=
     \tfrac{2}{3} + 2(1-y) - 2{(1-y)}^2 G(y) + \log(\mu^2/m_\pi^2),
\end{align}
$y\equiv 4m_\pi^2/q^2$, and
\begin{equation}
    G(y) =
    \begin{cases}
    \frac{1}{2\sqrt{1-y}}\left(
     \log\frac{1+\sqrt{1-y}}{1-\sqrt{1-y}} - i\pi \right) & \mbox{for $y<1$,}  \\[2.5ex]
     -\frac{1}{\sqrt{y-1}}\,\
     \arctan\left(1/\sqrt{y-1}\right) & \mbox{for $y>1$.}
     \end{cases}
\end{equation}
We normalize results from~\cite{Jegerlehner:2011ti} at $\mu=m_\pi$.

We are particularly interested in the corrected photon propagator from this theory,
since that is what enters $g-2$. Solving the
Lippmann-Schwinger equation gives:
\begin{equation}
\begin{split}
    G_{\gamma\gamma} =
    &\frac{1}{q^2\left(1 + \Sigma_{\gamma\gamma}^{(1)}\right)} \,+ \\
    &\frac{{\left(e/g + \Sigma_{\gamma\rho}^{(1)}\right)}^2}{
    q^2\left(1 + \Sigma_{\rho\rho}^{(1)}
    - {\left(e/g + \Sigma_{\gamma\rho}^{(1)}\right)}^2\right)
    - m_{0\rho}^2
    },
\end{split}
\end{equation}
where we have dropped a factor of $1 + \Sigma_{\gamma\gamma}^{(1)}$ in
the denominator of the last term since it enters only in order~$e^4$.

This propagator has poles at $q^2=0$ and at the (renormalized) $\rho$ mass. We can
set the coupling constants from the behavior near the $\rho$ pole, where
\begin{equation}
    G_{\gamma\gamma}(q) \to  \frac{e^2 f_\rho^2 }{2 m_\rho^2}\,
    \frac{1}{q^2
    - m_\rho^2 + im_\rho\Gamma_\rho}.
\end{equation}
Here $f_\rho$ is the $\rho$'s decay constant, and $\Gamma_\rho$ is its width.
Comparing these two expressions we find that:
\begin{align}
    m_\rho^2 - im_\rho\Gamma_\rho &= m_{0\rho}^2 \left(1
    - g_{\rho\pi\pi}^2\,\Sigma(m_\rho^2)\right)
\end{align}
and
\begin{align}
    \frac{f_\rho}{\sqrt{2}\,m_\rho} &=
    \frac{
    1/g_\rho + g_{\rho\pi\pi}\,\Sigma(m_\rho^2)
    }{
    1 + \tfrac{1}{2}\,g_{\rho\pi\pi}^2\,\Sigma(m_\rho^2)
    + \tfrac{1}{2}\,g_{\rho\pi\pi}^2\,m_\rho^2\,\Sigma^\prime(m_\rho^2)
    } \\
    &\approx
    \frac{1}{g_\rho}
    \left(
    1 + g_\rho\,g_{\rho\pi\pi}\,\Sigma(m_\rho^2)
    - \tfrac{1}{2}\,g_{\rho\pi\pi}^2\,\Sigma(m_\rho^2) \right. \nonumber \\
    &\quad\quad\quad\quad\left.
    -\tfrac{1}{2}\,g_{\rho\pi\pi}^2\,m_\rho^2\,\Sigma^\prime(m_\rho^2)
    \right)
\end{align}
up to QED corrections suppressed by~$\alpha_\mathrm{QED}$. Here
\begin{equation}
   48\pi^2\,q^2\, \Sigma^\prime(q^2) = 3 y - 1 - 3 y (1-y) G(y)
\end{equation}
where, again, $y\equiv 4m_\pi^2/q^2$.
Taking
\begin{align}
    m_\rho &= 0.775\,\mathrm{GeV} \quad
    \Gamma_\rho = 0.148\,\mathrm{GeV} \quad
    f_\rho \approx 0.21\,\mathrm{GeV} \nonumber \\
    m_\pi &= m_{\pi^+} = 0.1396\,\mathrm{GeV},
\end{align}
we find the bare parameters are:
\begin{equation}
    m_{0\rho} = 0.766\,\mathrm{GeV}
    \quad
    g_\rho = 5.4 \quad g_{\rho\pi\pi} = 6.0.
\end{equation}

\subsection{Contribution to $g-2$}
Returning to the photon propagator, we find that
\begin{equation}
    G_{\gamma\gamma} \to \frac{Z_\mathrm{had}}{q^2\left(1 - e^2\hat\Pi(q)\right)}
\end{equation}
near the photon pole, where
\begin{equation}
    Z_\mathrm{had} = \frac{1}{1 - e^2\Pi(0)},
\end{equation}
$\hat\Pi(q) \equiv \Pi(q) - \Pi(0)$, and
\begin{align}
    \Pi(q^2) &= -\Sigma(q^2) +
    \frac{q^2\left(1/g_\rho + g_{\rho\pi\pi}\,\Sigma(q^2)\right)^2}{
    q^2\left(1 + g_{\rho\pi\pi}^2\,\Sigma(q^2)\right) - m_{0\rho}^2
    }
\end{align}
This can be rewritten
\begin{align}
    \hat\Pi(q^2) &=  -\hat \Sigma(q^2) \nonumber \\
    &+ \frac{ \hat f^2}{2 \hat m^2}\,
    \frac{q^2\left(1 + g_\rho g_{\rho\pi\pi} \,\hat \Sigma(q^2) \right)^2}%
    {q^2 \left(1+ g^2_{\rho\pi\pi} \,\hat \Sigma(q^2)\right)-\hat m^2}
    \label{eq:PIhat-rho-pi}
\end{align}
where $\hat \Sigma(q^2) \equiv \mathrm{Re}\,\Sigma(q^2) - \Sigma(0)$,
\begin{align}
    \hat m^2 &\equiv m_0^2 \left( 1 - g_{\rho\pi\pi}^2\, \Sigma(0)
    \right) \\
    &= m_\rho^2 \left( 1 + g_{\rho\pi\pi}^2\,\hat \Sigma(m_\rho^2)
    \right),
\end{align}
and
\begin{align}
    \frac{\hat f}{\hat m} &\equiv \frac{\sqrt2}{g_\rho}
    \left(
    1 + g_\rho\, g_{\rho\pi\pi} \,\Sigma(0) - \tfrac{1}{2}\,g_{\rho\pi\pi}^2\, \Sigma(0)
    \right) \\
    &\approx \frac{f_\rho}{m_\rho}
    \left(
    1 - \tfrac{1}{2}\,g^2_{\rho\pi\pi} \,\hat \Sigma(m_\rho^2)
    + \tfrac{1}{2}\,g_{\rho\pi\pi}^2\, m_\rho^2 \,\hat \Sigma^\prime(m_\rho^2)
    \right)
\end{align}
are all independent of the ultraviolet regulator.
We approximated $g_\rho\to g_{\rho\pi\pi}$ in the last line above, to
simplify the result. Values for $\hat f$ and $\hat m$ equal those
for $f_\rho$ and $m_\rho$ to within a few percent.

To compute the contribution to $g-2$ from $\hat\Pi(q)$, we Taylor expand
and switch to Euclidean momenta ($q^2\to -q^2_E$):
\begin{equation}
    \hat\Pi(-q^2_E) \equiv \sum_{j=1}^\infty q^{2j}_E\,\Pi_j,
\end{equation}
where $\Pi_j = \Pi_j^{(\pi\pi)} + \Pi_j^{(\rho)}$, corresponding to the
first and second terms in Eq.~(\ref{eq:PIhat-rho-pi}), respectively.

To leading order,
\begin{align}
    \label{eq:pipi-moment}
    \Pi^{(\pi\pi)}_j &= \frac{{(-1)}^{j+1}}{8\pi^2\,m_\pi^{2j}}\,
    \frac{(j+1)!\, (j-1)!}{(2j+3)!}, \\
    \label{eq:rho-moment}
    \Pi^{(\rho)}_j &= \frac{{(-1)}^{j+1} f_\rho^2}{2 m_\rho^{2j+2}} + \mathcal{O}(g_{\rho\pi\pi}^2)
\end{align}
Substituting these results into our formalism for $g-2$,
with $m_\pi=m_{\pi^+}=0.13957$,
gives the leading contributions from $\pi\pi$ loops and from the $\rho$:
\begin{align}
    a_\mu^{(\pi\pi)} &= 71 \times 10^{-10} \\
    a_\mu^{(\rho)} &= 425 \times 10^{-10}  + \mathcal{O}(g_{\rho\pi\pi}^2)
\end{align}
This shows that the $\rho$ by itself accounts for about~71\% of
the total vacuum polarization contribution to~$a_\mu$, with
$\pi\pi$~interactions adding another~12\%.

\begin{table*}
\caption{Pion masses for different tastes, and the corresponding
finite-volume plus staggered-pion corrections to be added to the
Taylor coefficients~$\Pi_j$ for each configuration (as given
in Table~II).
The pion masses are based upon results in~\cite{Bazavov:2012uw},
using our definition of the lattice spacing. The Taylor
coefficients include an extra~10\% uncertainty,
beyond that due to uncertainties in the pion masses,
to account for uncalculated and partially calculated higher-order terms in
chiral perturbation theory.
}
\label{tab:stagg-pipi}
    \begin{ruledtabular}\begin{tabular}{cllllllllll}
Set & \ctitle{$m_\pi(\xi_5)$} & \ctitle{$m_\pi(\xi_{5\mu})$} & \ctitle{$m_\pi(\xi_{\mu\nu})$} & \ctitle{$m_\pi(\xi_\mu)$} & \ctitle{$m_\pi(1)$} & \ctitle{$\delta\Pi_1$} & \ctitle{$\delta\Pi_2$} & \ctitle{$\delta\Pi_3$} & \ctitle{$\delta\Pi_4$}  \\
\hline
1 & $0.302\hfill(2)$ & $0.362\hfill(3)$ & $0.407\hfill(4)$ & $0.451\hfill(5)$ & $0.485\hfill(19)$ & $0.0012\hfill(1)$ & $-0.0050\hfill(5)$ & $0.014\hfill(1)$ & $-0.034\hfill(4)$       \\
2 & $0.216\hfill(1)$ & $0.294\hfill(3)$ & $0.348\hfill(4)$ & $0.399\hfill(6)$ & $0.438\hfill(23)$ & $0.0028\hfill(3)$ & $-0.0160\hfill(16)$ & $0.063\hfill(7)$ & $-0.220\hfill(24)$     \\
3 & $0.133\hfill(1)$ & $0.240\hfill(3)$ & $0.304\hfill(5)$ & $0.362\hfill(7)$ & $0.405\hfill(26)$ & $0.0094\hfill(9)$ & $-0.0836\hfill(86)$ & $0.588\hfill(62)$ & $-4.320\hfill(472)$   \\
\hline
4 & $0.301\hfill(2)$ & $0.334\hfill(2)$ & $0.360\hfill(3)$ & $0.390\hfill(4)$ & $0.413\hfill(9)$ & $0.0008\hfill(1)$ & $-0.0038\hfill(4)$ & $0.012\hfill(1)$ & $-0.029\hfill(3)$     \\
5 & $0.218\hfill(1)$ & $0.262\hfill(2)$ & $0.295\hfill(3)$ & $0.331\hfill(4)$ & $0.359\hfill(11)$ & $0.0025\hfill(2)$ & $-0.0141\hfill(15)$ & $0.056\hfill(6)$ & $-0.196\hfill(22)$  \\
6 & $0.217\hfill(1)$ & $0.261\hfill(2)$ & $0.294\hfill(3)$ & $0.331\hfill(4)$ & $0.358\hfill(11)$ & $0.0022\hfill(2)$ & $-0.0131\hfill(13)$ & $0.054\hfill(6)$ & $-0.196\hfill(22)$  \\
7 & $0.216\hfill(1)$ & $0.261\hfill(2)$ & $0.294\hfill(3)$ & $0.330\hfill(4)$ & $0.358\hfill(11)$ & $0.0021\hfill(2)$ & $-0.0125\hfill(13)$ & $0.052\hfill(6)$ & $-0.191\hfill(21)$  \\
8 & $0.133\hfill(1)$ & $0.197\hfill(2)$ & $0.240\hfill(4)$ & $0.284\hfill(5)$ & $0.316\hfill(13)$ & $0.0081\hfill(8)$ & $-0.0771\hfill(79)$ & $0.571\hfill(60)$ & $-4.340\hfill(474)$ \\
\hline
9 & $0.308\hfill(2)$ & $0.319\hfill(2)$ & $0.328\hfill(2)$ & $0.337\hfill(2)$ & $0.345\hfill(4)$ & $0.0005\hfill(1)$ & $-0.0026\hfill(3)$ & $0.008\hfill(1)$ & $-0.021\hfill(2)$    \\
10 & $0.219\hfill(1)$ & $0.235\hfill(1)$ & $0.247\hfill(2)$ & $0.259\hfill(3)$ & $0.270\hfill(5)$ & $0.0013\hfill(1)$ & $-0.0084\hfill(9)$ & $0.038\hfill(4)$ & $-0.148\hfill(16)$  \\
\end{tabular}\end{ruledtabular}
\end{table*}

\subsection{Lattice Corrections}
\label{appendix:effth}
Lattice simulations modify the low-energy
analysis given above in two ways: 1) the lattice volume is finite;
and 2) pion-loop results are
averaged over several tastes of pion, each with a different mass.
The second of these is peculiar to formalisms, like HISQ, that use
staggered quarks. These effects are largest in the
$\pi\pi$~vacuum polarization function.
To correct for these simulation artifacts, we
reexamine the $\pi\pi$~contribution to
$\tfrac{1}{3}\sum_i\Pi^{(\pi\pi)}_{ii}(q^2_E)$
in continuum Euclidean QCD:
\begin{align}
 \frac{4}{3} \int \frac{d^4k}{(2\pi)^4}\,
    \frac{k^2 - k_0^2}{(k^2 + m^2_a)(k^2 + 2k_0 q_E + q^2_E + m^2_b)}
    \\
\quad = \frac{4}{3} \int \! \frac{d^3\mathbf{k}}{(2\pi)^3 \,2E_aE_b} \,
    \frac{(E_a+E_b)\,\mathbf{k}^2}{q^2_E + (E_a + E_b)^2}
\label{eq:pipi-cont}
\end{align}
where $q_\mu = q^\mu = (q_E,0,0,0)$, $E_i=\sqrt{\mathbf{k}^2+m_i^2}$,
and normally $m_a=m_b=m_\pi$. This implies
that  the $\pi\pi$~vacuum polarization function
used in the previous section is given by:
\begin{align}
    &-\hat\Sigma(-q^2_E,m_a,m_a)
    \equiv
    \nonumber \\
    &\frac{4q^2_E}{3}\! \int \! \frac{d^3\mathbf{k}}{(2\pi)^3 2E_aE_b} \,
    \frac{\mathbf{k}^2}{(E_a+E_b)^3 (q^2_E + (E_a + E_b)^2)}.
    \label{eq:pipi-latt}
\end{align}
The Taylor coefficients $\Pi_j^{(\pi\pi)}$ derived in the previous section
are the coefficients of $q^{2j}_E$ in the expansion of this expression
when~$m_a=m_b=m_\pi$.

We correct for the finite spatial volume ($L^3$) of the lattice by replacing
\begin{equation} \label{eq:pipi-int}
    \int \frac{d^3\mathbf{k}}{(2\pi)^3} \to
    \frac{1}{L^3}\sum_{k_x=-\infty}^\infty \sum_{k_y=-\infty}^\infty
    \sum_{k_z=-\infty}^\infty
\end{equation}
where the sums are over discrete momenta $k=2\pi n/L$ for all integer~$n$,
positive and non-positive.
(We correct our fits separately for the finite temporal length of the lattice,
which is, in any case, 1.5--3~times longer than in spatial
directions and so effectively infinite.)
We ignore the effect of the finite lattice
spacing since the contributions of interest  are all ultraviolet finite
(and quite infrared). 

The second modification concerns the pion masses
in the vacuum polarization, and is specific to staggered-quark actions like the
HISQ~action we use. In our simulations we use vector currents~$J_\mu$
that are local, which means that they carry taste~$\xi_\mu$.
(We use the notation of~\cite{HISQ_PRD}, which discusses quark doubling
and taste symmetry at length, especially in Appendices~A--D.)
Taste conservation means that the pion pairs must carry the
same total taste as the current but there are
several different taste pairings that accomplish this.
A current with total taste $\xi_\mu$ can couple to pion pairs carrying tastes:
\begin{enumerate}
    \item $\xi_5\oplus\xi_{\mu5}$ (2 combinations);
    \item $\xi_{\nu5}\oplus\xi_{\rho\sigma}$ where~$\mu$, $\nu$, $\rho$,
    and~$\sigma$ are all different (6 combinations);
    \item $\xi_{\rho\sigma}\oplus\xi_{\nu}$ where~$\rho=\mu\ne\nu=\sigma$
    (6 combinations);
    \item $\xi_{\mu}\oplus 1$ (2 combinations).
\end{enumerate}
The total contribution is the average over these 16~possibilities.
We estimate the $\pi\pi$~contribution to the
vacuum polarization in our simulations by averaging over the contributions
Eq.~(\ref{eq:pipi-latt}) from each pairing of pion tastes, with $m_a$ and $m_b$ set
to the masses of the two pions.
We use pion masses for different tastes derived from
MILC's results in~\cite{Bazavov:2012uw} (see Table~\ref{tab:stagg-pipi}).

In Table~\ref{tab:stagg-pipi} we list corrections~$\delta\Pi_j$
for the moments from
each of our configuration sets. We add these to the Monte Carlo results
in order to correct for effects due to the finite volume and
pion-mass taste splittings. We estimate these corrections by
approximating Eq.~(\ref{eq:PIhat-rho-pi}) with
\begin{align}
    \hat\Pi(&-q^2_E,f_\rho,m_\rho,m_\pi) =
    -\hat \Sigma(-q^2_E,m_\pi,m_\pi) \nonumber \\
    &+ \frac{ f_\rho^2}{2 m^2_\rho}\,
    \frac{q^2_E\left(1 +
    g_\rho g_{\rho\pi\pi} \,\hat \Sigma(-q^2_E,m_\pi,m_\pi) \right)^2}%
    {q^2_E \left(1+ g^2_{\rho\pi\pi} \,
    \hat \Sigma(-q^2_E,m_\pi,m_\pi)\right) + m^2_\rho}
    \label{eq:pi-hat-e}
\end{align}
where $\hat\Sigma$ is the $\pi^+\pi^-$ vacuum polarization
function from Eq.~(\ref{eq:pipi-latt}), and
we have replaced $\hat f$ and $\hat m$ by~$f_\rho$ and~$m_\rho$,
respectively.
To obtain the
correction for a
given configuration set, we first evaluate
this continuum vacuum polarization function
using the (Goldstone)~$m_\pi$, $m_\rho$,
and~$f_\rho$ obtained from the configuration set (Table~\ref{tab:params}), and
then we subtract from it the same quantity but with
\begin{equation}
    \hat \Sigma(-q^2_E,m_\pi,m_\pi) \to \frac{1}{16}\sum_{\xi_a,\xi_b}
    \hat\Sigma_V(-q^2_E,m_\pi(\xi_a),m_\pi(\xi_b))
\end{equation}
where $\hat\Sigma_V$ is evaluated for the finite volume of the configuration
(Eq.~(\ref{eq:pipi-int})), and averaged over the staggered-pion taste
combinations~$\xi_a\oplus\xi_b$ listed above. The corrections~$\delta\Pi_j$
are the Taylor coefficients of this difference between continuum and
finite-volume/staggered-pion vacuum polarizations.

\begin{figure}
\centering
\includegraphics{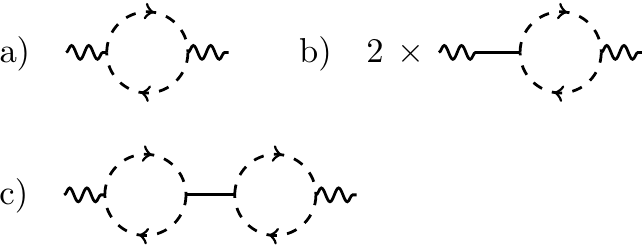}
\caption{Leading
diagrams from the $\rho$ effective field theory that
correspond (to leading order in $q^2/m_\rho^2)$
to the diagrams in Fig.~\ref{fig:chpth} from the standard
chiral theory:
a) leading-order $\pi^+\pi^-$ vacuum polarization;
b) correction for the pion's charge radius from
$\gamma\to\rho\to\pi\pi$;
c) correction for $\pi\pi$~scattering correction from
$\pi\pi\to\rho\to\pi\pi$. Dashed and solid lines represent
pions and rhos, respectively.
}
\label{fig:rhoth}
\end{figure}

The contribution to~$a_\mu$
from the first term in Eq.~(\ref{eq:pi-hat-e}) is roughly
five times larger than that from the second term,
and has the opposite sign. This is for our simulation results with physical
pion masses and the intermediate lattice spacing (set~8).
The largest contributions come mainly from the terms
\begin{align}
    &-\hat\Sigma(-q^2_E,m_\pi,m_\pi) \times \Big( 1 -
    g_\rho g_{\rho\pi\pi} \frac{f_\rho^2}{m_\rho^2}\,\frac{q_E^2}{m_\rho^2}
    \Big) \\
    &\quad\quad =  -\hat\Sigma(-q^2_E,m_\pi,m_\pi) \times
    \Big(1 - \frac{\langle r_\pi^2 \rangle q_E^2}{3}\Big)
\end{align}
in Eq.~(\ref{eq:pi-hat-e}) (Figs.~\ref{fig:rhoth}a and~\ref{fig:rhoth}b),
where $r_\pi$~is the charge radius of the pion. They contribute
corrections to $a_\mu$ of~$50\times10^{-10}$ and~$-13\times10^{-10}$,
respectively. Further
$(q^2/m_\rho^2)^n$ corrections to the $\gamma$-$\pi\pi$ vertex
contribute~$3\times10^{-10}$. The other $q^2_E/m_\rho^2$
correction in Eq.~(\ref{eq:pi-hat-e})
is from $\pi\pi$~scattering (Fig.~\ref{fig:rhoth}c):
\begin{equation}
    \frac{f_\rho^2}{2m_\rho^2}\,\frac{q_E^2}{m_\rho^2}\,\Big(
    g_\rho g_{\rho\pi\pi}\hat\Sigma(-q_E^2,m_\pi,\pi)\Big)^2.
\end{equation}
This should be small because it is
second order in $g_\rho g_{\rho\pi\pi}\hat\Sigma$; in fact, it
contributes less than~$0.5\times10^{-10}$. The total correction from
all contributions (to all orders)
is $41\times10^{-10}$ for set~8\,---\,chiral perturbation
theory converges relatively rapidly here.

We add an extra 10\%~uncertainty to each correction~$\delta\Pi_j$ to
account for missing contributions suppressed by~$m_s/\Lambda$, due to
tadpole and other renormalizations of the leading vacuum polarization.
This uncertainty also accounts for corrections of order~$(q^2/\Lambda)^2$
and higher that are only partially included by our analysis.

\begin{figure}
\centering
\includegraphics[width=0.48\textwidth]{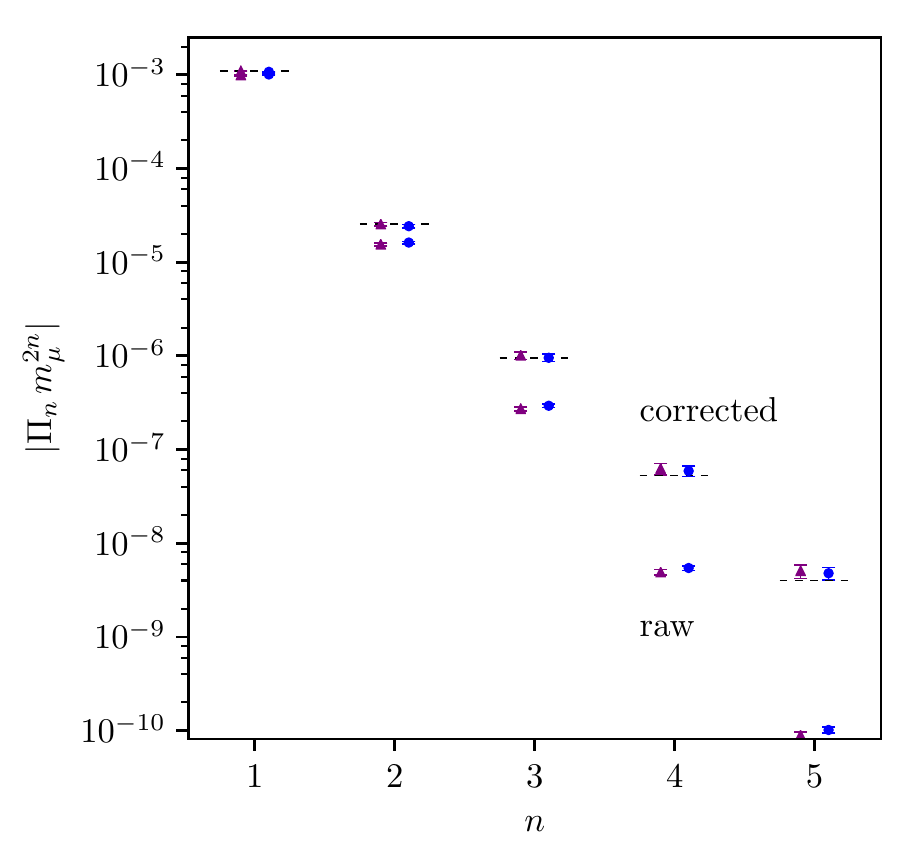}
\caption{Contributions to the hadronic vacuum
polarization~$\hat\Pi(q^2)$ at~$q^2=-m_\mu^2$
coming from individual Taylor coefficients~$\Pi_n$ with $n=1\ldots5$.
Results are show for corrected (above) and
uncorrected (``raw'', below) coefficients coming
from our lattice QCD simulations with
physical sea-quark masses (sets 3 and 8).
The corrected coefficents include both corrections
described in Section~\ref{sec:correction-amu}:
1)~adding $\delta \Pi_n$ from Table~\ref{tab:stagg-pipi}; and 2)~replacing
the pion mass from the simulation with the physical pion mass in the
leading $\pi\pi$~loop. To compare with experiment,
we add contributions
from $s$~and $c$~quarks~\cite{Chakraborty:2014mwa}
to both the raw and corrected moments,
neglecting their contribution to the $n=5$~moment (which is negligible).
The dashed lines are results derived from $e^+e^-$ data:
see the ``data direct'' column in Table~I of~\cite{Benayoun:2016krn}.
The error estimates on the
lattice results do not include contributions due
to electromagnetic, isospin-violating, and disconnected contributions;
(estimated to be around 2\% for the $n=1$~moment).}
\label{fig:taylorcoef}
\end{figure}

\begin{figure}
\centering
\includegraphics[width=0.48\textwidth]{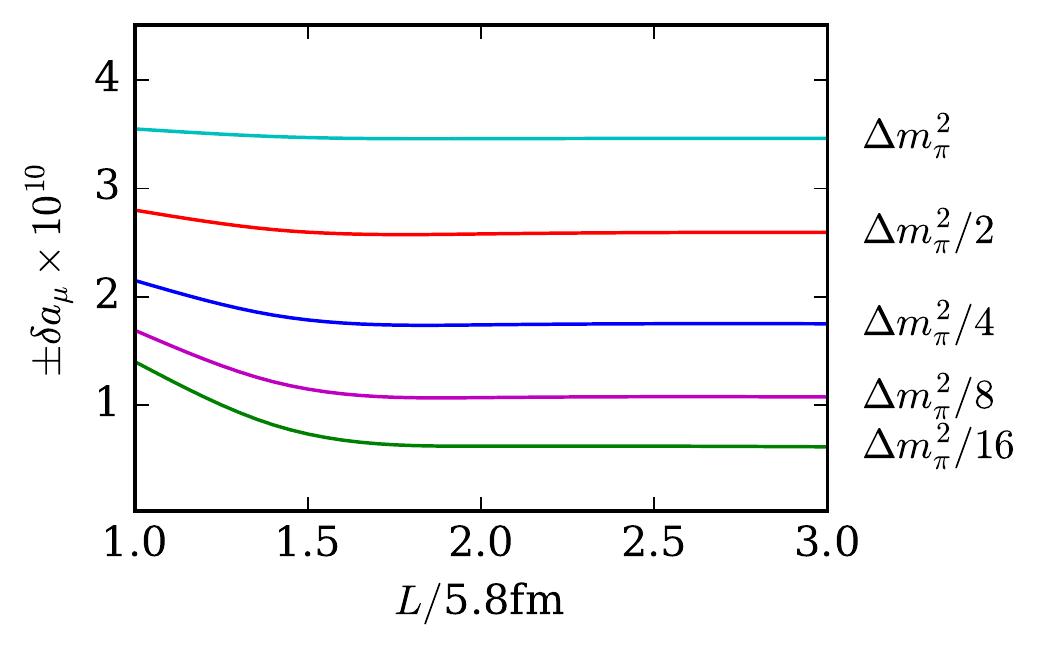}
\caption{
Uncertainty in $a_{\mu}^{\mathrm{HVP,LO}}$ due to
finite-volume and staggered-pion effects as a function
of the average taste-splitting $\Delta m_\pi^2$ between
pions and the spatial size~$L$ of the lattice at the physical
value of $m_{\pi^+}$ (140 MeV).
Here the line marked $\Delta m_\pi^2$ refers to the splittings for
configuration set~8 in Table~\ref{tab:stagg-pipi} for which
$L=5.8$\,fm. The splittings decrease slightly faster
than $a^2$ as the lattice spacing decreases, so the other lines shown
correspond to conservative uncertainties at lattice spacings of approximately
0.09\,fm, 0.06\,fm, 0.045\,fm and 0.03\,fm. The
uncertainties are estimated to be~$1/10$ of the correction.
}
\label{fig:fvol}
\end{figure}

The taste structure of the $\pi\pi$~vacuum polarization matters
because its contribution to~$a_\mu$ is quite sensitive to the
pion mass (see Eq.~(\ref{eq:pipi-moment}))
and pions of different taste differ in mass. Taste-changing
interactions normally lead to small corrections that extrapolate
smoothly to zero, like~$\alpha_s(\pi/a)\,a^2$, as the lattice
spacing vanishes. This does not work for the $\pi\pi$~vacuum
polarization with physical pions, however,
because its moments are non-analytic in
$m_\pi$ (Eq.~(\ref{eq:pipi-moment})) and the taste-changing
effects are comparable to the (physical) pion mass. This is why
we use chiral perturbation theory to remove the effects of the staggered
pion masses in the $\pi\pi$~vacuum polarization.
There are other effects from taste-changing but we only need
correct contributions that are non-analytic in~$m_\pi$ (and large enough
to matter); all other effects will extrapolate away as we take
the lattice spacing to zero. The $a$-independence of our final results
is evidence that we have handled these corrections properly.

As noted in the main text, the largest corrections ($7$\%) are for our
lightest pion masses. Corrections for our heaviest pions are
about an order of magnitude smaller, and therefore
negligible compared to other errors.
The corrections are also negligible for $s$-quark vacuum polarization,
as discussed in our previous paper~\cite{Chakraborty:2014mwa}.

We tested our finite-volume analysis by analyzing simulations with three
different volumes for our intermediate lattice spacing and a pion mass
of about 220\,MeV (configuration sets~5--7). The raw data show variations
between the three volumes of~3.1(1.3)\%. Our corrections,
from finite-volume/staggered-pion-masses and $\rho$-mass rescaling,
reduce this variation
by an order of magnitude; see Figure~\ref{fig:amu}.
This is a non-trivial test of our corrections.

We also tested our finite-volume/staggered-pion corrections
by comparing results for individual
Taylor coefficients with experiment.
Fig~\ref{fig:taylorcoef} shows the corrected lattice results, combined
with $s$~and $c$~quark contributions from~\cite{Chakraborty:2014mwa},
together with
results based on data from $e^+e^-$~annihilation~\cite{Benayoun:2016krn}. The
agreement is strong evidence that our estimates for these corrections
are reliable. Note that the corrections for moments with~$n\ge3$
are larger than the uncorrected results from the simulation.
These large-$n$ moments have negligible impact on~$a_\mu$~($\le0.5$\%), but
they provide sensitive tests of our corrections.

The $\delta\Pi_j$ are almost entirely due to the staggering of the
pion masses. Normally one would expect larger finite-volume errors,
but here the average pion mass
appearing in any $\pi\pi$ vacuum polarization contribution
is larger than the physical pion mass because of the staggering.
This strongly suppresses finite-volume effects.
Fig.~\ref{fig:fvol} shows how the uncertainty from this
correction depends upon the taste-splittings between pions~$\Delta m_\pi^2$
and the spatial size~$L$ of the lattice.
Lines are drawn for varying $\Delta m_\pi^2$ at
physical pion mass starting from coarse set~8.
The uncertainty shown in the figure for the largest
$\Delta m_{\pi}^2$ is somewhat smaller than the uncertainty
that we use for configuration set~8 because the pion mass
on that ensemble is smaller than the physical pion mass.